\documentclass[journal]{IEEEtran}
\usepackage[T1]{fontenc}
\usepackage[latin9]{inputenc}
\usepackage{float}
\usepackage{mathrsfs}
\usepackage{amsmath}
\usepackage{amssymb}
\usepackage{stmaryrd}
\usepackage{graphicx}
\usepackage[bookmarks=true,bookmarksnumbered=true,bookmarksopen=true,bookmarksopenlevel=1,
 breaklinks=false,pdfborder={0 0 0},pdfborderstyle={},backref=false,colorlinks=false]
 {hyperref}
\hypersetup{pdftitle={Your Title},
 pdfauthor={Your Name},
 pdfborderstyle=,pdfpagelayout=OneColumn,pdfnewwindow=true,pdfstartview=XYZ,plainpages=false}

\makeatletter

\floatstyle{ruled}
\newfloat{algorithm}{tbp}{loa}
\providecommand{\algorithmname}{Algorithm}
\floatname{algorithm}{\protect\algorithmname}

\usepackage{algorithmic}
\usepackage{cuted}

\makeatother

\begin{document}
\title{Antenna Coding Optimization for Pixel Antenna Empowered MIMO Wireless
Power Transfer}
\author{Yijun Chen,~\IEEEmembership{Graduate~Student~Member,~IEEE,} Shanpu
Shen,~\IEEEmembership{Senior Member,~IEEE,} Tianrui Qiao,~\IEEEmembership{Member,~IEEE,}
Hongyu Li,~\IEEEmembership{Graduate~Student~Member,~IEEE,} Kai-Kit
Wong,~\IEEEmembership{Fellow,~IEEE,} and Ross Murch,~\IEEEmembership{Fellow,~IEEE}
\thanks{This work was funded by the Hong Kong Research Grants Council for
the General Research Fund (GRF) grant 16208124 and the Science and
Technology Development Fund, Macau SAR (File/Project no. 001/2024/SKL).
(\textit{Corresponding author: Shanpu Shen}.)} \thanks{Yijun Chen, Tianrui Qiao, and Ross Murch are with the Department of
Electronic and Computer Engineering, The Hong Kong University of Science
and Technology, Clear Water Bay, Kowloon, Hong Kong.} \thanks{Shanpu Shen is with the State Key Laboratory of Internet of Things
for Smart City and Department of Electrical and Computer Engineering,
University of Macau, Macau, China (e-mail: shanpushen@um.edu.mo).}
\thanks{Hongyu Li is with the Internet of Things Thrust, The Hong Kong University
of Science and Technology (Guangzhou), Guangzhou 511400, China.} \thanks{Kai-Kit Wong is with the Department of Electronic and Electrical Engineering,
University College London, Torrington Place, WC1E 7JE, United Kingdom
and Yonsei Frontier Lab, Yonsei University, Seoul, Korea.} }

\maketitle
\newcounter{MYtempeqncnt}
\begin{abstract}
We investigate antenna coding utilizing pixel antennas as a new degree
of freedom for enhancing multiple-input multiple-output (MIMO) wireless
power transfer (WPT) systems. The objective is to enhance the output
direct current (DC) power under RF combining and DC combining schemes
by jointly exploiting gains from antenna coding, beamforming, and
rectenna nonlinearity. We first propose the MIMO WPT system model
with binary and continuous antenna coding using the beamspace channel
model and formulate the joint antenna coding and beamforming optimization
using a nonlinear rectenna model. We propose two efficient closed-form
successive convex approximation algorithms to efficiently optimize
the beamforming. To further reduce the computational complexity, we
propose codebook-based antenna coding designs for output DC power
maximization based on K-means clustering. Results show that the proposed
pixel antenna empowered MIMO WPT system with binary antenna coding
increases output DC power by more than 15 dB compared with conventional
systems with fixed antenna configuration. With continuous antenna
coding, the performance improves another 6 dB. Moreover, the proposed
codebook design outperforms previous designs by up to 40\% and shows
good performance with reduced computational complexity. Overall, the
significant improvement in output DC power verifies the potential
of leveraging antenna coding utilizing pixel antennas to enhance WPT
systems.
\end{abstract}


\section{Introduction}

\IEEEPARstart{W}{ith} the rapid development of wireless sensor networks
(WSN) and the Internet of Things (IoT) in the past decades, there
has been an explosive increase in devices including wearable devices,
RFIDs, and biomedical implants. However, energizing a large number
of devices poses a critical challenge, as battery replacement or recharging
becomes prohibitive and unsustainable. Wireless power transfer (WPT),
where the radio frequency (RF) energy transmitted from a dedicated
energy transmitter (ET), without any wires, is received and rectified
into direct current (DC) energy at an energy receiver (ER), provides
a potentially convenient and reliable solution to energize a large
number of devices \cite{ClerckxFoundations}.

The\textbf{ }main challenge of WPT is to maximize the end-to-end power
transfer efficiency, i.e. maximizing the output DC power for a given
transmit power. To this end, one approach is efficient rectenna design
with enhanced performance in aspects of antenna gain, bandwidth, and
impedance matching \cite{wagih2020rectennas,ullah2022review}. In
addition, compact multiport rectennas \cite{shen2019ambient,shen2017dual,shen2017multiport,shen2020directional}
have been developed to maximize the output DC power, but lacks a system-level
consideration. Another approach is efficient signal design for WPT,
including beamforming, waveform, modulation, and power allocation
designs \cite{clerckx2018fundamentals,zeng2017communications}. Previous
works \cite{clerckx2016waveform,shen2020beamforming,shen2021joint,huang2017large,moghadam2017waveform}
have studied the design of multi-sine WPT waveforms that adapt to
channel conditions to maximize the output DC power. In \cite{shen2020beamforming}
and \cite{shen2021joint}, systematic designs of multi-antenna WPT
have been proposed, where the beamforming gain, frequency-selectivity
and rectenna nonlinearity are jointly leveraged under DC combining
and RF combining schemes. Nevertheless, current WPT systems are mainly
based on antennas with fixed configuration, while overlooking the
potential of reconfigurable antennas to further boost the output DC
power.

Pixel antennas are a promising highly reconfigurable antenna technology
\cite{cetiner2004multifunctional,zhang2022highly,zhang2022low} that
can be incorporated into WPT systems to achieve a breakthrough in
boosting output DC power. Pixel antennas consist of discretized sub-wavelength
elements referred to as pixels, which can be connected together with
RF switches. By controlling the states of RF switches, pixel antennas
support a wide range of reconfigurability in antenna characteristics
such as radiation pattern \cite{zhang2022highly,zhang2022low,rao2022novel},
operating frequency \cite{chiu2012frequency,6678331}, and polarization
\cite{zheng2024design}. Closely related to pixel antennas, there
is also an emerging technology (introduced in 2020), termed the fluid
antenna system (FAS) \cite{wong2020performance,wong2020fluid}. The
concept of FAS is based on dynamic adjustment of antenna positions
within a linear region to obtain better channel conditions and enhance
wireless communication \cite{psomas2023diversity,zhu2023modeling}.
FAS can be implemented by mechanical movement using liquids or by
pixel antennas to mimic the position adjustment through optimizing
RF switch configuration \cite{zhang2024novel,wong2025reconfigurable}.
Leveraging the antenna position adjustable property, FAS has been
applied to different applications including fluid antenna multiple
access (FAMA) \cite{wong2022fast}, wireless powered NOMA systems
\cite{ghadi2025physical}, simultaneous wireless information and power
transfer (SWIPT) \cite{zhou2025fluid}, and FAMA-assisted integrated
data and energy transfer \cite{lin2025fluid}.

While the development of FAS has demonstrated the benefits of reconfigurable
antennas in wireless systems, to further leverage the potential of
pixel antennas and generalize the radiation pattern reconfigurability,
a novel technology denoted as antenna coding has been recently proposed
in \cite{11202491}. In antenna coding binary variables called antenna
coders are utilized to represent the states of RF switches. It has
been demonstrated that antenna coding technology can enhance the channel
gain of single-input single-output (SISO) systems and the channel
capacity of multiple-input multiple-output (MIMO) systems \cite{11202491}.
In addition, it has also been shown to enhance the sum rate in multi-user
MISO systems, where pixel antennas are deployed at the user side,
through jointly optimizing antenna coding and transmit precoding \cite{li2025antenna}.
Considering the limitation of WPT systems based on antennas with fixed
configuration and the significant potential of pixel antennas in designing
and enhancing wireless systems, it is useful to investigate pixel
antennas with antenna coding optimization in WPT systems. This approach
promises to overcome the limitation of low power transfer efficiency
in WPT, which remains an open challenging problem.

In this work, we aim to jointly optimize antenna coding with beamforming
and combining for pixel antenna MIMO WPT systems to maximize the output
DC power. We further broaden and deepen the investigation \cite{Chen2512:Antenna}
by considering continuous antenna coding, lower complexity algorithms,
RF combining scenarios, and codebook-based antenna coding designs.
The contributions of this work are summarized as follows.

\emph{First}, we propose the pixel antenna MIMO WPT system with both
binary and continuous antenna coding to greatly enhance the output
DC power to overcome the limitation of low power transfer efficiency.
To that end, we introduce the MIMO beamspace channel model, which
can demonstrate the pattern reconfigurability utilizing pixel antennas
and reduce channel estimation overhead in MIMO WPT systems.

\emph{Second}, we formulate and solve the MIMO WPT output DC power
maximization problem for DC combining schemes, where the transmit
beamforming and antenna coding designs are jointly considered and
alternatively optimized. Specifically, to handle the complex non-convex
transmit beamforming design, we derive closed-form solutions based
on a successive convex approximation (SCA) method, which is more efficient
than previous geometric programming with semi-definite relaxation
(SDR) methods \cite{shen2020beamforming}. We also propose both binary
and continuous antenna coding designs, by using Successive Exhaustive
Boolean Optimization (SEBO) and the quasi-Newton method, respectively.

\emph{Third}, we formulate and solve the MIMO WPT DC power maximization
problem for RF combining schemes. With the alternating optimization
approach, we propose both binary and continuous antenna coding designs
and closed-form solution for beamforming designs. In particular, we
also propose an efficient SCA algorithm for the analog receive beamforming
design, which is a practical and low-complexity RF combining scheme
by utilizing phase shifters.

\emph{Fourth}, we propose an efficient codebook design based on K-means
clustering to reduce the computational complexity of binary antenna
coding design, which considers the MIMO WPT configuration and the
objective of output DC power maximization problems. After the offline
codebook training, the antenna coder for each pixel antenna is online
deployed by searching the codebook.

\emph{Fifth}, we evaluate the average output DC power of the proposed
pixel antenna empowered MIMO WPT system with antenna coding. Compared
with conventional systems with fixed antenna configuration, using
binary antenna coding can significantly enhance the average output
DC power by more than 15 dB. Moreover, using continuous antenna coding
brings another 6 dB gain compared with binary antenna coding. The
proposed codebook-based antenna coding design also outperforms the
random codebook and existing design \cite{11202491} by up to 40\%,
providing good performance but with much lower computational complexity.

\emph{Organization}: Section II introduces a beamspace channel model
and MIMO WPT system model with binary and continuous antenna coding.
Section III presents the joint antenna coding and beamforming design
with DC combining scheme. Section IV presents the joint antenna coding
and beamforming design with RF combining scheme. Section V presents
the codebook-based antenna coding design. Section VI provides performance
evaluations for the proposed pixel antenna empowered MIMO WPT system.
Section VII concludes this work.

\emph{Notation}: Bold lowercase and uppercase letters represent vectors
and matrices, respectively. A symbol without bold font denotes a scalar.
$\mathcal{E}\{\cdot\}$, $\mathfrak{R}\{\cdot\},$ $\mathbb{R}$ and
$\mathbb{C}$ represent expectation, real part, real and complex number
sets, respectively. $[\mathbf{a}]_{i}$, $\mathbf{a}^{*}$, and $\|\mathbf{a}\|$
denote the $i$th element, conjugate, and $l_{2}$-norm of a vector
$\mathbf{a}$, respectively. $\mathbf{A}^{\mathrm{\mathit{T}}}$,
$\mathbf{A}^{H}$, $[\mathbf{A}]_{i,:}$, $[\mathbf{A}]_{:,i}$, and
$[\mathbf{A}]_{i,j}$ denote the transpose, conjugate transpose, $i$th
row, $i$th column, and $(i,j)$th element of a matrix $\mathbf{A}$,
respectively. $\cup$ denotes the union of sets. $\mathcal{CN}(\mathbf{0},\mathbf{\Sigma})$
represents the circularly symmetric complex Gaussian distribution
with zero mean and covariance matrix $\mathbf{\Sigma}$. diag$(a_{1},...,a_{N})$
denotes a diagonal matrix with entries $a_{1},...,a_{N}$. blkdiag$(\mathbf{a}_{1},...,\mathbf{a}_{N})$
denotes a block diagonal matrix formed by $\mathbf{a}_{1},...,\mathbf{a}_{N}$.
$j$ is the imaginary unit. $\mathbf{I}$ denotes the identity matrix.

\section{MIMO WPT System Utilizing Pixel Antennas}

The basic principle of MIMO WPT systems utilizing pixel antennas is
introduced in this section, including the concept of antenna coding,
beamspace channel model, and MIMO WPT system model.

\subsection{Pixel Antennas and Antenna Coding}

\begin{figure}[t]
\begin{centering}
\includegraphics[width=8.5cm]{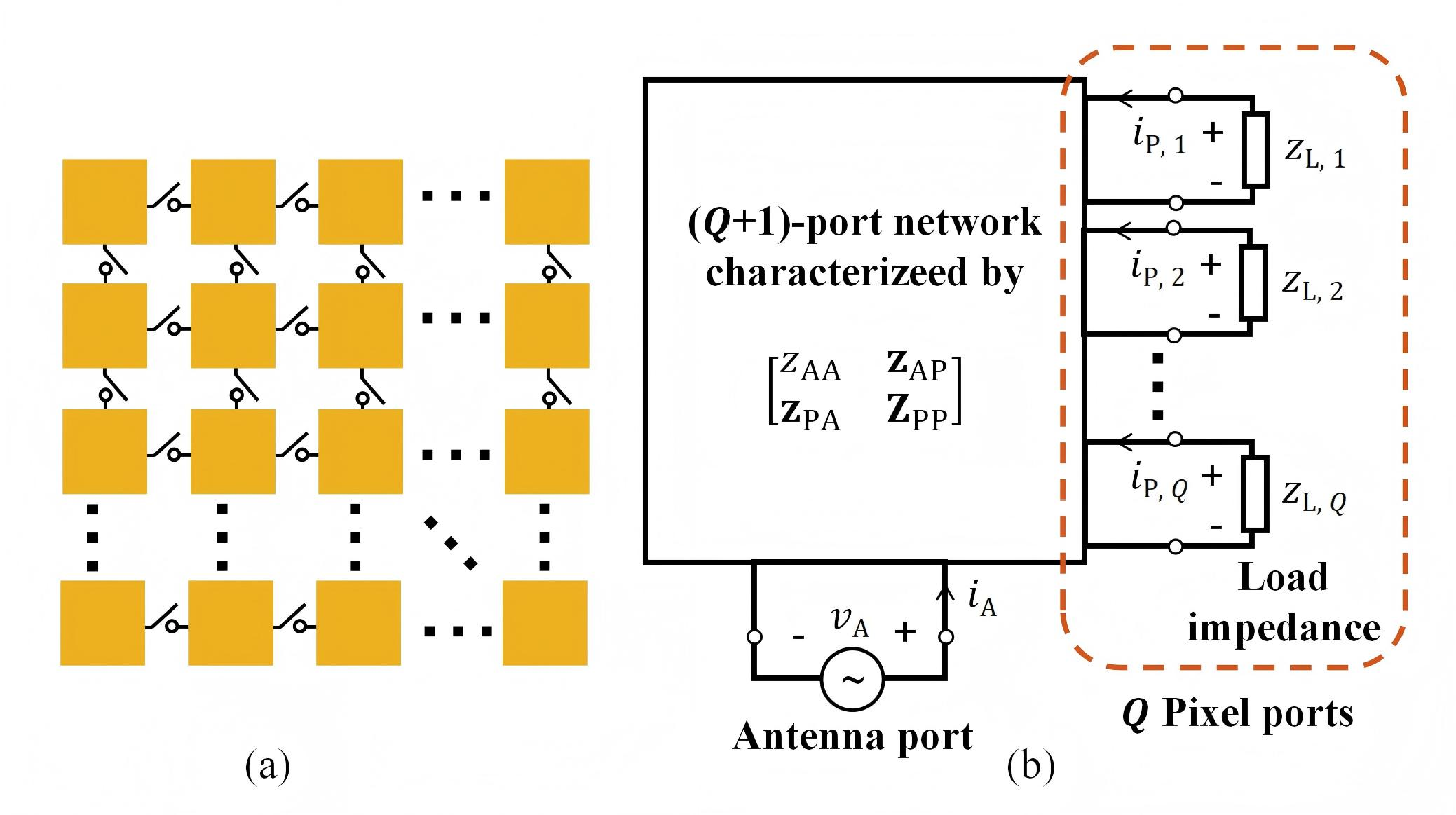}
\par\end{centering}
\caption{Schematics of (a) pixel antenna and (b) multiport network model.}\label{fig:Schematic-of-(a)}
\end{figure}

We first briefly introduce the pixel antenna model and the corresponding
antenna coding technology. As shown in Fig. \ref{fig:Schematic-of-(a)}(a),
the pixel antenna is based on a discretized radiation surface consisting
of sub-wavelength pixels. Adjacent pixels are connected through $Q$
RF switches or variable reactive loads, which can be flexibly adjusted
to reconfigure the antenna characteristics. According to multiport
network theory \cite{6678331}, the pixel antenna can be accurately
modeled as a $(Q+1)$-port network with one antenna port and $Q$
pixel ports as shown in Fig. \ref{fig:Schematic-of-(a)}(b). We characterize
the $(Q+1)$-port network by its impedance matrix $\mathbf{Z}=[z_{\mathrm{AA}},\mathbf{z}_{\mathrm{AP}};\mathbf{z}_{\mathrm{PA}},\mathbf{Z}_{\mathrm{PP}}]\in\mathbb{C}^{(Q+1)\times(Q+1)}$,
where $z_{\mathrm{AA}}\in\mathbb{C}$ and $\mathbf{Z}_{\mathrm{PP}}\in\mathbb{C}^{Q\times Q}$
denote the self-impedance (matrix) for the antenna port and pixel
ports respectively, $\mathbf{z}_{\mathrm{AP}}\in\mathbb{C}^{1\times Q}$
and $\mathbf{z}_{\mathrm{PA}}\in\mathbb{C}^{Q\times1}$ represent
the trans-impedance between the antenna port and pixel ports with
$\mathbf{z}_{\mathrm{AP}}=\mathbf{z}_{\mathrm{PA}}^{T}$. The voltages
and currents of the $(Q+1)$-port network are therefore related by
\begin{equation}
\begin{bmatrix}v_{\mathrm{A}}\\
\boldsymbol{\mathbf{v}}_{\mathrm{P}}
\end{bmatrix}=\begin{bmatrix}z_{\mathrm{AA}} & \mathbf{z}_{\mathrm{AP}}\\
\mathbf{z}_{\mathrm{PA}} & \boldsymbol{\mathbf{Z}}_{\mathrm{PP}}
\end{bmatrix}\begin{bmatrix}i_{\mathrm{A}}\\
\boldsymbol{\mathbf{i}}_{\mathrm{P}}
\end{bmatrix},\label{eq: Zi=00003Dv}
\end{equation}
where $v_{\mathrm{A}}\in\mathbb{C}$ and $\boldsymbol{\mathbf{v}}_{\mathrm{P}}=[v_{\mathrm{P},1},...,v_{\mathrm{P},Q}]^{T}\in\mathbb{C}^{Q\times1}$
denote the voltages at the antenna port and pixel ports, and the currents
at the antenna port and pixel ports are $i_{\mathrm{A}}\in\mathbb{C}$
and $\boldsymbol{\mathbf{i}}_{\mathrm{P}}=[i_{\mathrm{P},1},...,i_{\mathrm{P},Q}]^{T}\in\mathbb{C}^{Q\times1}$,
respectively.

Depending on whether adjacent pixels are connected by either RF switches
or variable reactive loads, we achieve binary antenna coding or continuous
antenna coding, as detailed below.

\subsubsection{Binary Antenna Coding}

Leveraging RF switches, adjacent pixels in the pixel antenna can be
reconfigured to be either connected or unconnected. This enables the
antenna topology to be reconfigured to create the desired antenna
characteristics. Accordingly, the $q$th RF switch can be modeled
as load impedance $\mathrm{\mathit{z}}_{\mathrm{L},q}\in\mathbb{C}$,
which is either short-circuit or open-circuit depending on the switch
on (connected) or off (unconnected) state. Therefore, we can use a
binary variable $b_{q}\in\{0,1\}$ to characterize the switch state,
i.e. $b_{q}=0$ for switch on state and $b_{q}=1$ for switch off
state, so that we have
\begin{equation}
\mathrm{\mathit{z}}_{\mathrm{L},q}=\begin{cases}
0, & \mathrm{if}\,b_{q}=0,\\
\infty, & \mathrm{if}\,b_{q}=1,
\end{cases}\label{eq: Zlq, binary}
\end{equation}
where the infinity for the switch off state, i.e. the open-circuit
load impedance, can be numerically approximated by a very large value
$jx_{\mathrm{oc}}$, e.g. $x_{\mathrm{oc}}=10^{9}$. Thus, we can
rewrite \eqref{eq: Zlq, binary} as 
\begin{equation}
\mathrm{\mathit{z}}_{\mathrm{L},q}=jx_{\mathrm{oc}}b_{q}.\label{mapping}
\end{equation}
We collect $b_{q}$ $\forall q$ into a vector $\mathrm{\mathbf{b}=}\mathrm{[\mathit{\mathrm{\mathit{b_{\mathrm{1}},...,b_{\mathit{Q}}}}}]^{\mathit{T}}\in\mathbb{R}^{\mathit{Q\times\mathrm{1}}}}$
to denote all switch states, which is denoted here as the antenna
coder.

\subsubsection{Continuous Antenna Coding}

To enhance the reconfigurablity of the pixel antenna, we can extend
the binary antenna coding to continuous antenna coding by replacing
the RF switches with variable reactive loads such as varactors. Similarly,
the $q$th variable reactive load can be modeled as load impedance
$\mathrm{\mathit{z}}_{\mathrm{L},q}$, written as 
\begin{equation}
\mathrm{\mathit{z}}_{\mathrm{L},q}=jx_{\mathrm{L},q},\label{mapping-1}
\end{equation}
where $x_{\mathrm{L},q}\in\mathbb{R}$ is the load reactance for the
$q$th variable reactive load. The load reactance $x_{\mathrm{L},q}$
can be positive/negative infinity in theory, while numerically it
can be expressed as
\begin{equation}
\left|x_{\mathrm{L},q}\right|=x_{\mathrm{oc}}b_{q},\:b_{q}\in\left[0,1\right],\label{mapping-1-1}
\end{equation}
where the antenna coder $b_{q}$ $\forall q$ can be continuous in
the range $[0,1]$ and the binary antenna coding with the open/short-circuit
load impedance \eqref{mapping} is a special case.

For both binary and continuous antenna coding, we can group $\mathrm{\mathit{z}}_{\mathrm{L},q}$
$\forall q$ into a diagonal load impedance matrix $\mathbf{Z}_{\mathrm{L}}(\mathbf{b})=\mathrm{diag}(\mathrm{\mathit{z}}_{\mathrm{L},1},...,\mathrm{\mathit{z}}_{\mathrm{L},Q})\in\mathbb{C}^{\mathit{Q\times Q}}$,
which is coded by $\mathbf{b}$, so that $\boldsymbol{\mathbf{v}}_{\mathrm{P}}$
and $\boldsymbol{\mathbf{i}}_{\mathrm{P}}$ can be related by 
\begin{equation}
\boldsymbol{\mathbf{v}}_{\mathrm{P}}=-\mathbf{Z}_{\mathrm{L}}(\boldsymbol{\mathrm{b}})\boldsymbol{\mathbf{i}}_{\mathrm{P}}.\label{eq: load v i}
\end{equation}

Substituting \eqref{eq: load v i} into \eqref{eq: Zi=00003Dv}, we
can obtain the currents at pixel ports $\mathbf{i}_{\mathrm{P}}(\mathbf{b})$
with excitation $i_{\mathrm{A}}$ as 
\begin{equation}
\mathbf{i}_{\mathrm{P}}(\mathbf{b})=-(\mathbf{Z}_{\mathrm{PP}}+\mathbf{Z}_{\mathrm{L}}(\mathbf{b}))^{-1}\mathbf{z}_{\mathrm{PA}}i_{\mathrm{A}}.\label{eq: current ip}
\end{equation}
We collect the currents at all ports into a vector $\mathbf{i}(\mathbf{b})=[i_{\mathrm{A}};\mathbf{i}_{\mathrm{P}}(\mathbf{b})]\in\mathbb{C}^{(Q+1)\times1}$.

The radiation pattern $\mathbf{e}(\mathbf{b})\in\mathbb{C}^{2K\times1}$
generated by the pixel antenna is given by
\begin{equation}
\mathrm{\mathbf{e}(\mathbf{b})=\mathbf{\mathbf{e}}_{\mathrm{A}}\mathit{i}_{\mathrm{A}}+\sum_{\mathit{q}=1}^{\mathit{Q}}\mathbf{\mathbf{e}}_{\mathrm{\mathrm{P,\mathit{q}}}}\mathit{i}_{P,\mathit{q}}(\mathbf{b})=\mathbf{E}_{\mathrm{oc}}\mathbf{i}(\mathbf{b})},
\end{equation}
where $\mathrm{\mathbf{e}_{\mathrm{A}}\in\mathbb{C}^{\mathit{\mathrm{2\mathit{K}\times1}}}}$
and $\mathbf{e}_{\mathrm{\mathrm{P}},q}\in\mathbb{C}^{2K\times1}$
are the radiation patterns of the antenna port and $q$th pixel port
(with $\theta$ and $\phi$ polarization components over $K$ sampled
spatial angles) excited by unit current when other ports are open-circuit,
respectively, and $\mathbf{E_{\mathrm{oc}}=\mathrm{[\mathbf{e}_{\mathrm{A}},\mathbf{e}_{\mathrm{\mathrm{P},1}},...,\mathbf{e}_{\mathrm{\mathrm{P},\mathit{Q}}}]}}\in\mathbb{C}^{\mathrm{\mathit{\mathrm{2}K\times(Q+\mathrm{1})}}}$
is the open-circuit radiation pattern matrix for all ports. For the
binary and continuous antenna coding cases, the antenna coders are
optimized among $\mathrm{2^{\mathit{Q}}}$ discrete combinations and
in a continuous space, respectively. With the optimized antenna coders,
the antenna characteristics such as radiation patterns can be flexibly
reconfigured to adapt to the channel, providing extra degrees of freedom
to design and enhance WPT systems.

\subsection{MIMO Beamspace Channel Model}

\begin{figure}
\begin{centering}
\includegraphics[width=8.5cm]{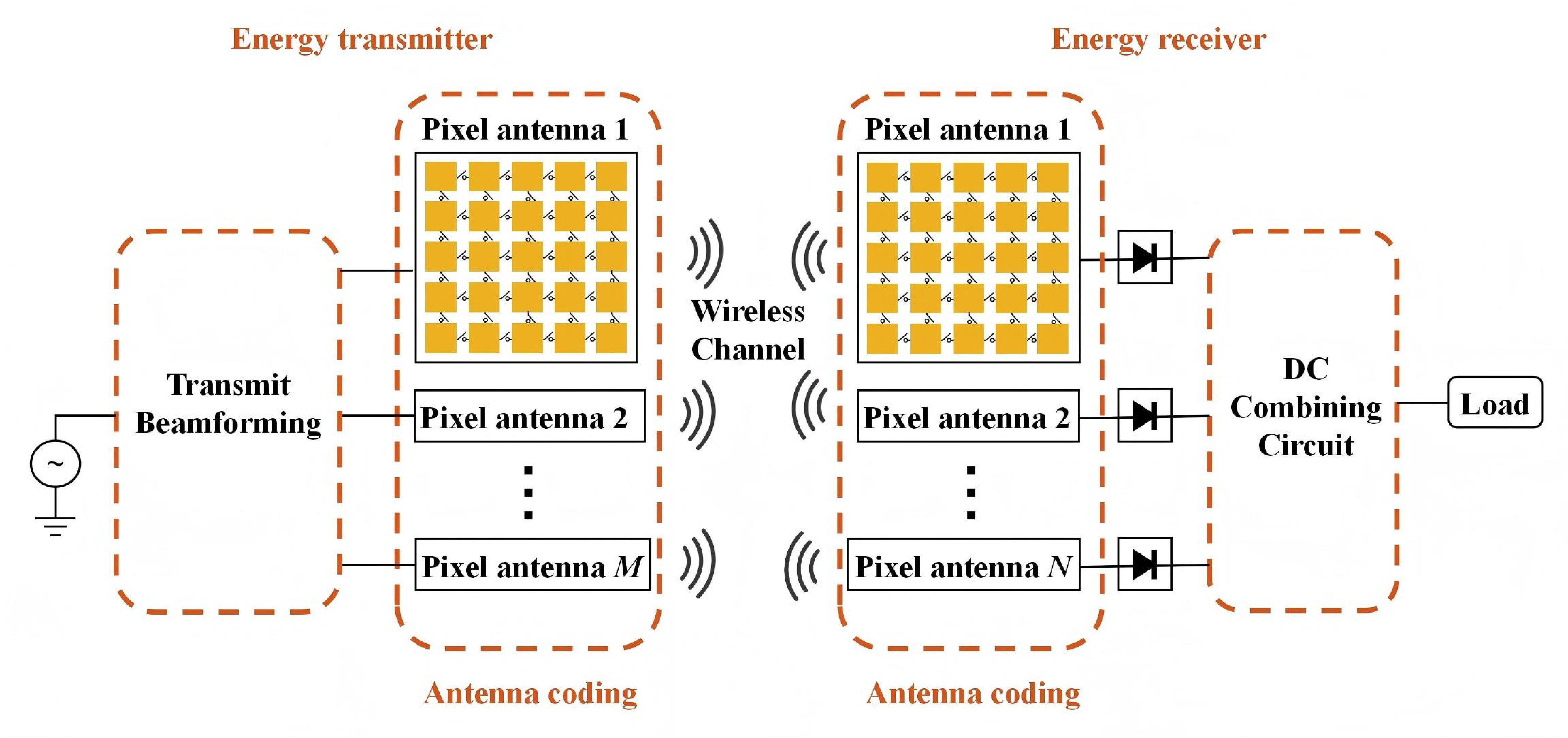}
\par\end{centering}
\centering{}\caption{Schematic of the proposed MIMO WPT system using pixel antennas with
DC combining.}\label{fig:Schematic-of-the}
\end{figure}

\begin{figure}[t]
\centering
\begin{centering}
\includegraphics[width=8.5cm]{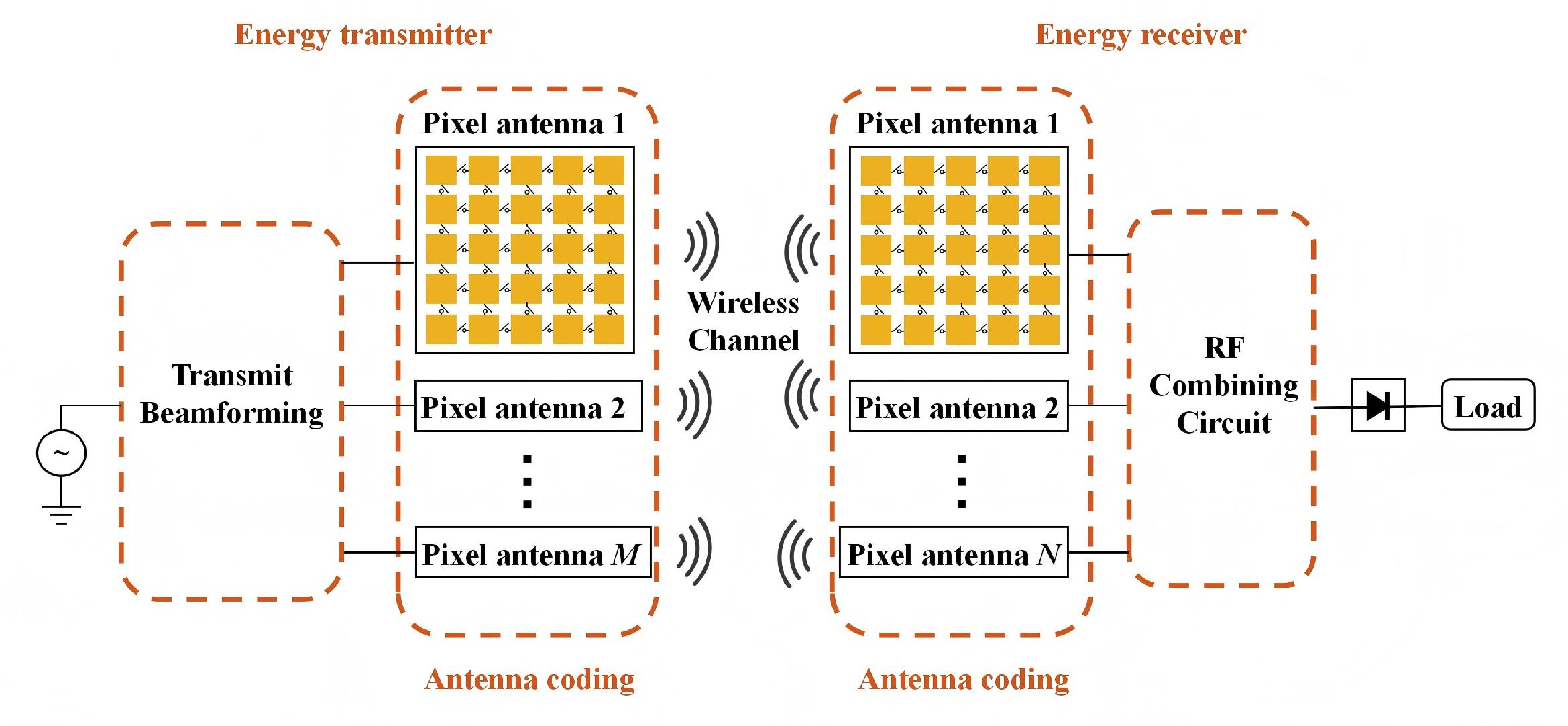}
\par\end{centering}
\centering{}\caption{Schematic of the proposed MIMO WPT system using pixel antennas with
RF combining.}\label{fig:Schematic-of-the-1}
\end{figure}

We consider a MIMO pixel antenna system with $M$ pixel antennas at
the transmitter (each coded by $\mathbf{b}_{\mathrm{T,\mathit{m}}},m=1,2,...,M$)
and $N$ pixel antennas at the receiver (each coded by $\mathbf{b}_{\mathbf{\mathrm{R,\mathit{n}}}},n=1,2,...,N$)
as shown in Fig. \ref{fig:Schematic-of-the} and Fig. \ref{fig:Schematic-of-the-1},
where we assume all pixel antennas are identical and spatially separated.
Given the currents of the transmit and receive pixel antennas, $\mathbf{i}_{\mathrm{T},m}(\mathbf{b}_{\mathrm{T},m})\in\mathbb{C}^{(Q+1)\times1}\:\forall m$
and $\mathbf{i}_{\mathrm{R},n}(\mathbf{b}_{\mathrm{R},n})\in\mathbb{C}^{(Q+1)\times1}\:\forall n$,
the radiation pattern of each transmit and receive pixel antenna can
be respectively written as
\begin{align}
\mathbf{e}_{\mathrm{T},m}(\mathbf{b}_{\mathrm{T},m}) & =\mathrm{\mathbf{E}_{\mathbf{\mathrm{oc,T,\mathit{m}}}}\mathbf{i}_{\mathrm{T},m}(\mathbf{b}_{\mathrm{T},m})},\\
\mathbf{e}_{\mathrm{R},n}(\mathbf{b}_{\mathrm{R},n}) & =\mathbf{E}_{\mathbf{\mathrm{oc,R,\mathit{n}}}}\mathbf{i}_{\mathrm{R},n}(\mathbf{b}_{\mathrm{R},n}),
\end{align}
where $\mathbf{E}_{\mathbf{\mathrm{oc,T,\mathit{m}}}}\in\mathbb{C}^{\mathrm{\mathit{\mathrm{2}K\times(Q+\mathrm{1})}}}$
and $\mathbf{E}_{\mathbf{\mathrm{oc,R,\mathit{n}}}}\in\mathbb{C}^{\mathrm{\mathit{\mathrm{2}K\times(Q+\mathrm{1})}}}$
are the open-circuit radiation pattern matrices of the $m$th transmit
and $n$th receive pixel antenna, respectively.

Using the beamspace channel representation \cite{johnson1992array},
the channel for the MIMO pixel antenna system can be written as 
\begin{equation}
\mathbf{H}(\mathbf{B}_{\mathrm{T}},\mathbf{B}_{\mathrm{R}})=\mathbf{E_{\mathrm{R}}^{\mathit{T}}\mathrm{(\mathbf{B}_{\mathrm{R}})}H_{\mathrm{V}}\mathbf{E_{\mathrm{T}}\mathrm{(}\mathbf{B}_{\mathrm{T}}\mathrm{)}}},\label{BS}
\end{equation}
where $\mathbf{E_{\mathrm{T}}\mathrm{(}\mathbf{B}_{\mathrm{T}}\mathrm{)}}=[\mathbf{e}_{\mathrm{T,1}}(\mathbf{b}_{\mathrm{T},1}),...,\mathbf{e}_{\mathrm{T,\mathit{M}}}(\mathbf{b}_{\mathrm{T},M})]\in\mathbb{C}^{\mathrm{\mathit{\mathrm{2}K\times M}}}$
and $\mathbf{E_{\mathrm{R}}\mathrm{(\mathbf{B}_{\mathrm{R}})}}=[\mathbf{e}_{\mathrm{R,1}}(\mathbf{b}_{\mathrm{R},1}),...,\mathbf{e}_{\mathrm{R,\mathit{N}}}(\mathbf{b}_{\mathrm{R},N})]\in\mathbb{C}^{\mathrm{\mathit{\mathrm{2}K\times N}}}$
collect normalized radiation patterns where $\|\mathbf{e}_{\mathrm{T},m}(\mathbf{b}_{\mathrm{T},m})\|=1\:\forall m$
and $\|\mathbf{e}_{\mathrm{R},n}(\mathbf{b}_{\mathrm{R},n})\|=1\:\forall n$
are for channel normalization, $\mathbf{B}_{\mathrm{T}}=\left[\mathbf{b}_{\mathrm{T},1},\ldots,\mathbf{b}_{\mathrm{T},M}\right]\in\mathbb{R}^{Q\times M}$
and $\mathbf{B}_{\mathrm{R}}=\left[\mathbf{b}_{\mathrm{R},1},\ldots,\mathbf{b}_{\mathrm{R},N}\right]\in\mathbb{R}^{Q\times N}$
are antenna coder matrices for transmit and receive pixel antennas
respectively, and $\mathbf{H}_{\mathrm{V}}\in\mathbb{C}^{\mathrm{\mathit{\mathrm{2}K\times\mathrm{2}K}}}$
is the virtual channel matrix collecting channel gains between all
pairs of angle of departure (AoD) and angle of arrival (AoA) among
the dual polarizations and $K$ sampled spatial angles.

To demonstrate the additional degrees of freedom provided by pixel
antennas, we first perform singular value decomposition (SVD) of the
open-circuit radiation pattern matrices $\mathbf{E}_{\mathbf{\mathrm{oc,T,\mathit{m}}}}\:\forall\mathit{m}$
and $\mathbf{E}_{\mathbf{\mathrm{oc,R,\mathit{n}}}}\:\forall n$ of
all the transmit and receive pixel antennas. We assume that all pixel
antennas are identical, so that the open-circuit radiation patterns
of different pixel antennas have the same shape. Thus, $\mathbf{E}_{\mathbf{\mathrm{oc,T,\mathit{m}}}}\:\forall m$
and $\mathbf{E}_{\mathbf{\mathrm{oc,R,\mathit{n}}}}\:\forall n$ share
the same rank, which can be denoted as $N_{\mathrm{eff}}$. Specifically,
using SVD, we have that 
\begin{align}
\mathrm{\mathbf{E}_{\mathbf{\mathrm{oc,T,\mathit{m}}}}} & =\mathbf{U}_{\mathrm{T},\mathit{m}}\mathbf{S}_{\mathrm{T},\mathit{m}}\mathbf{V}_{\mathrm{T},\mathit{m}}^{\mathit{H}},\\
\mathbf{E}_{\mathbf{\mathrm{oc,R,\mathit{n}}}} & =\mathbf{U}_{\mathrm{R},\mathit{n}}\mathbf{S}_{\mathrm{R},\mathit{n}}\mathbf{V}_{\mathrm{R},\mathit{n}}^{\mathit{H}},
\end{align}
where $\mathbf{S}_{\mathrm{T},\mathit{m}},\mathbf{S}_{\mathrm{R},\mathit{n}}\in\mathbb{R}^{N_{\mathrm{eff}}\times N_{\mathrm{eff}}}$
are diagonal matrices containing $N_{\mathrm{eff}}$ non-zero singular
values, $\mathrm{\mathbf{U}_{\mathrm{T},\mathit{m}}}\in\mathbb{C}^{\mathrm{\mathit{\mathrm{2}K\times N_{\mathrm{eff}}}}}$,
$\mathbf{U}_{\mathrm{R},\mathit{n}}\in\mathbb{C}^{\mathrm{\mathit{\mathrm{2}K\times N_{\mathrm{eff}}}}}$,
$\mathbf{V}_{\mathrm{T},\mathit{m}}\in\mathbb{C}^{\mathrm{\mathit{(Q+\mathrm{1})\times N_{\mathrm{eff}}}}}$,
and $\mathbf{V}_{\mathrm{R},\mathit{n}}\in\mathbb{C}^{\mathrm{\mathit{(Q+\mathrm{1})\times N_{\mathrm{eff}}}}}$
are semi-unitary matrices of the $m$th transmit and $n$th receive
pixel antenna respectively. $N_{\mathrm{eff}}$ is also referred to
as the effective aerial degrees-of-freedom (EADoF) \cite{han2023using},
which indicates the number of orthogonal radiation patterns the pixel
antenna can provide. The $N_{\mathrm{eff}}$ columns of each $\mathbf{U}_{\mathrm{T},\mathit{m}}\:\forall m$
and $\mathbf{U}_{\mathrm{R},\mathit{n}}\:\forall n$ can be utilized
as the orthogonal basis patterns to construct the beamspace. Therefore,
the MIMO pixel antenna system can be equivalently viewed as a conventional
MIMO system with $\mathrm{\mathit{N_{\mathrm{T}}=MN_{\mathrm{eff}}}}$
and $\mathit{N_{\mathrm{R}}=NN_{\mathrm{eff}}}$ spatially separated
transmit and receive antennas.

Accordingly, we can define the pattern coders for each transmit and
receive pixel antennas as 
\begin{align}
\mathbf{w}_{\mathrm{T,\mathit{m}}}\text{(}\mathbf{b}_{\mathrm{T,\mathit{m}}}\mathrm{)} & =\mathbf{\mathbf{S}_{\mathrm{T},\mathit{m}}\mathbf{V}_{\mathrm{T},\mathit{m}}^{\mathit{H}}}\mathbf{i}_{\mathrm{T},m}(\mathbf{b}_{\mathrm{T},m}),\label{eq: wt}\\
\mathbf{w}_{\mathrm{R,\mathit{n}}}\text{(}\mathbf{b}_{\mathrm{R,\mathit{n}}}\mathrm{)} & =\mathbf{S}_{\mathrm{R},\mathit{n}}\mathbf{V}_{\mathrm{R},\mathit{n}}^{\mathit{T}}\mathbf{i}_{\mathrm{R},n}^{*}(\mathbf{b}_{\mathrm{R},n}),\label{eq:wr}
\end{align}
which satisfy $\left\Vert \mathbf{w}_{\mathrm{T,\mathit{m}}}\text{(}\mathbf{b}_{\mathrm{T,\mathit{m}}}\mathrm{)}\right\Vert =1$
and $\left\Vert \mathbf{w}_{\mathrm{R,\mathit{n}}}\text{(}\mathbf{b}_{\mathrm{R,\mathit{n}}}\mathrm{)}\right\Vert =1.$
With the pattern coder, we can rewrite the radiation pattern of the
$m$th transmit and $n$th receive pixel antenna as 
\begin{align}
\mathbf{e}_{\mathrm{T,\mathit{m}}}\text{(}\mathbf{b}_{\mathrm{T,\mathit{m}}}\mathrm{)} & =\mathbf{U_{\mathrm{T,\mathit{m}}}}\mathbf{w}_{\mathrm{T,\mathit{m}}}\text{(}\mathbf{b}_{\mathrm{T,\mathit{m}}}\mathrm{)},\label{eq: et=00003DUtwt}\\
\mathbf{e}_{\mathrm{R,\mathit{m}}}\text{(}\mathbf{b}_{\mathrm{R,\mathit{m}}}\mathrm{)} & =\mathbf{U_{\mathrm{R,\mathit{m}}}}\mathbf{w}_{\mathrm{R,\mathit{m}}}^{*}\text{(}\mathbf{b}_{\mathrm{R,\mathit{m}}}\mathrm{)},\label{eq: er=00003DUrwr}
\end{align}
which implies that the radiation pattern of the pixel antenna can
be synthesized by linearly coding the orthogonal basis patterns with
a pattern coder. Based on \eqref{eq: wt} and \eqref{eq:wr}, the
overall transmit and receive pattern coders are given as block-diagonal
matrices $\mathbf{W_{\mathrm{T}}\mathrm{(}B_{\mathrm{T}}\mathrm{)}}=\mathrm{blkdiag}(\mathbf{w_{\mathrm{T,1}}\mathrm{(}b_{\mathrm{T,1}}\mathrm{)}},...,\mathbf{w_{\mathrm{T,\mathit{M}}}\mathrm{(}b_{\mathrm{T,\mathit{M}}}\mathrm{)}})\in\mathbb{C}^{\mathrm{\mathit{N_{\mathrm{T}}\times M}}}$
and $\mathbf{W_{\mathrm{R}}\mathrm{(}B_{\mathrm{R}}\mathrm{)}}=\mathrm{blkdiag}(\mathbf{w_{\mathrm{R,1}}\mathrm{(}b_{\mathrm{R,1}}\mathrm{)}},...,\mathbf{w_{\mathrm{R,\mathit{N}}}\mathrm{(}b_{\mathrm{R,\mathit{N}}}\mathrm{)}})\in\mathbb{C}^{\mathrm{\mathit{N_{\mathrm{R}}\times N}}}$.
Utilizing the pattern coder definition to substitute \eqref{BS},
an equivalent beamspace channel model can be written as 
\begin{equation}
\mathbf{H}(\mathbf{B}_{\mathrm{T}},\mathbf{B}_{\mathrm{R}})=\mathbf{W_{\mathit{\mathrm{R}}}^{\mathit{H}}\mathrm{(}B_{\mathrm{R}}\mathrm{)}H_{\mathrm{C}}W_{\mathrm{T}}\mathrm{(}B_{\mathrm{T}}\mathrm{)}},\label{BS1-1}
\end{equation}
where $\mathbf{H}_{\mathrm{C}}\in\mathbb{C}^{\mathrm{\mathit{\mathrm{\mathit{N_{\mathrm{R}}}}\times\mathrm{\mathit{N_{\mathrm{T}}}}}}}$
denotes the channel between the $\mathrm{\mathit{N_{\mathrm{T}}}}$
and $\mathrm{\mathit{N_{\mathrm{R}}}}$ equivalent spatially separated
antennas, written as 
\begin{equation}
\mathbf{H_{\mathrm{C}}}=\mathbf{E_{\mathrm{bs,R}}^{\mathrm{\mathit{T}}}H_{\mathrm{V}}E_{\mathrm{bs,T}}},\label{eq:hc}
\end{equation}
where $\mathbf{E_{\mathrm{bs,T}}}=[\mathbf{U}_{\mathrm{T,1}},...,\mathbf{U}_{\mathrm{T,\mathit{M}}}]\in\mathbb{C}^{\mathrm{\mathit{\mathrm{2}K\times N_{\mathrm{T}}}}}$
and $\mathbf{E}_{\mathrm{bs,R}}=[\mathbf{U}_{\mathrm{R,1}},...,\mathbf{U}_{\mathrm{R,\mathit{N}}}]\in\mathbb{C}^{\mathrm{\mathit{\mathrm{2}K\times N_{\mathrm{R}}}}}$
collect the orthogonal basis patterns of spatially separated transmit
and receive pixel antennas, respectively. The formulation \eqref{eq:hc}
has the benefits of reducing channel estimation overhead and optimization
complexity as $\mathbf{H_{\mathrm{C}}}$ has a lower dimension than
the virtual channel matrix $\mathbf{H}_{\mathrm{V}}$. Considering
a rich scattering propagation environment with Rayleigh fading, we
can model $[\mathbf{H}_{\mathrm{C}}]_{i,j}\sim\mathcal{CN}(0,1)\,\forall i,j$
as independent and identically distributed (i.i.d.) complex Gaussian
random variables, which is consistent with the conventional $\mathrm{\mathit{N_{\mathrm{R}}}}\times\mathrm{\mathit{N_{\mathrm{T}}}}$
MIMO Rayleigh channel.

\subsection{MIMO WPT System Model}

For the proposed MIMO WPT system consisting of $M$ transmit pixel
antennas and $N$ receive pixel antennas, we denote the transmit beamformer
as $\mathbf{p}_{\mathrm{T}}=[p_{\mathrm{T},1},...,p_{\mathrm{T},M}]^{T}\in\mathbb{C}^{M\times1}$,
where $\mathrm{\mathit{p}_{\mathrm{T},\mathit{m}}\:\forall}m$ is
the beamforming weight at the $m$th transmit pixel antenna, and it
satisfies the power constraint $\frac{1}{2}\parallel\mathbf{p}_{\mathrm{T}}\parallel^{2}\leq P$,
where $P$ is the transmit power. Accordingly, the\textbf{ }transmit
signal $\mathbf{x}(t)\in\mathbb{C}^{M\times1}$ at time $t$ with
transmit beamforming can be written as 
\begin{equation}
\mathrm{\mathbf{x}(\mathit{t})=\Re\{\mathbf{p}_{\mathrm{T}}\mathit{e}^{\mathit{j\omega_{\mathrm{\mathrm{c}}}t}}\},}
\end{equation}
where $\omega_{\mathrm{c}}$ is the center frequency. The transmit
signal then propagates through the beamspace channel $\mathbf{H}(\mathbf{B}_{\mathrm{T}},\mathbf{B}_{\mathrm{R}})$.
Using \eqref{BS1-1}, the received signal $\mathbf{y}\mathrm{(\mathit{t})}\in\mathbb{C}^{N\times1}$
can be expressed as 
\begin{equation}
\mathrm{\mathit{\mathbf{y}}\mathrm{(\mathit{t})}=\Re\{\mathbf{W_{\mathit{\mathrm{R}}}^{\mathit{\mathrm{\mathit{H}}}}\mathrm{(}B_{\mathrm{R}}\mathrm{)}H_{\mathrm{C}}W_{\mathrm{T}}\mathrm{(}B_{\mathrm{T}}\mathrm{)}}\mathbf{p}_{\mathrm{T}}\mathit{e}^{\mathit{j\omega_{\mathrm{c}}t}}\}}.\label{eq:RF-1}
\end{equation}

We assume the channel matrix $\mathbf{H_{\mathrm{C}}}$ is perfectly
known to the transmitter and receiver for subsequent antenna coding
and beamforming optimization. The CSI can be accurately acquired by
channel estimation based on beamspace pilot transmission and feedback
\cite{zhang2022analog}.

\section{Antenna Coding Optimization for MIMO WPT with DC Combining}

In this section, we formulate and solve the antenna coding optimization
problem for MIMO WPT with DC combining (DCC) to maximize the output
DC power. An efficient algorithm to jointly optimize antenna coding
and transmit beamforming is also proposed.

For DCC scheme, as shown in Fig. \ref{fig:Schematic-of-the}, the
received signal at each receive pixel antenna is individually rectified.
We consider a nonlinear rectenna model in \cite{clerckx2016waveform,huang2017large},
so that the output DC voltage of the $n$th rectifier is given by
\begin{align}
v_{\mathrm{out},n} & \mathrm{=\sum_{\mathit{i\,\mathrm{even},\,i\geq\mathrm{2}}}^{\mathit{n_{\mathrm{0}}}}\beta_{\mathit{i}}\mathscr{\mathcal{E}}\left\{ \mathit{y_{n}}(\mathit{t}){}^{\mathit{i}}\right\} ,}\label{eq:vout}
\end{align}
where $n_{0}$ is the truncation order, e.g. $n_{0}=4$, $\mathrm{\beta_{\mathit{i}}=\frac{\mathit{R}_{ant}^{\mathit{i}/2}}{\mathit{i}!(\mathit{I}_{d}\mathit{v}_{t})^{\mathit{i}-1}}}$
is a constant with $\mathit{R}_{\mathrm{ant}}$, $\mathrm{\mathit{v}_{t}}$,
and $\mathit{I}_{\mathrm{d}}$ denoting the antenna impedance, thermal
voltage, and ideality factor, respectively, and $\mathscr{\mathcal{E}}\left\{ \mathit{y_{n}}(\mathit{t})^{\mathit{i}}\right\} $
is given by
\begin{equation}
\mathrm{\mathscr{\mathcal{E}}\left\{ \mathit{y_{n}}(\mathit{t}){}^{\mathit{i}}\right\} =}\zeta_{i}\left|[\mathbf{H}\left(\mathbf{B}_{\mathrm{T}},\mathbf{B}_{\mathrm{R}}\right)\mathbf{p}_{\mathrm{T}}]_{n}\right|^{i},\label{eq: average ith term}
\end{equation}
where $\mathrm{\zeta_{\mathit{i}}=\frac{1}{2\pi}\int_{0}^{2\pi}\sin^{\mathit{i}}\mathit{t}\textrm{d}\mathit{t}}$
and specifically $\zeta_{2}=\frac{1}{2}$, and $\zeta_{4}=\frac{3}{8}$.
The individually rectified DC power is then combined using circuits
such as a MIMO switching DC-DC converter \cite{liu2018dual}, which
yields the output DC power
\begin{equation}
\mathrm{\mathit{P}_{\mathrm{out}}^{\mathrm{DCC}}(\mathbf{p}_{\mathrm{T}},\mathbf{B}_{\mathrm{T}},\mathrm{\mathbf{B}_{\mathrm{R}}})=\sum_{\mathit{n}=1}^{\mathit{N}}\frac{\mathit{v_{\mathrm{out},n}^{\mathrm{2}}}}{\mathit{R}_{L}},}\label{eq:pdcc}
\end{equation}
where $\mathit{R}_{\mathrm{L}}$ denotes the load impedance of each
rectifier.

\subsection{Binary Antenna Coding Optimization with DCC}

Based on the nonlinear rectenna model, we aim to jointly optimize
binary antenna coding and transmit beamforming designs for maximizing
the output DC power, 

\begin{align}
\mathrm{\underset{\mathbf{p}_{T},\mathbf{B}_{T},\mathbf{B}_{R}}{\mathrm{max}}\;} & \mathrm{\mathit{P}_{\mathrm{out}}^{\mathrm{DCC}}(\mathbf{p}_{\mathrm{T}},\mathbf{B}_{\mathrm{T}},\mathrm{\mathbf{B}_{\mathrm{R}}})}\label{p1}\\
\mathrm{s.t.}\;\;\;\;\; & \frac{1}{2}\left\Vert \mathbf{p}_{\mathrm{T}}\right\Vert ^{2}\mathrm{\leq\mathit{P}},\\
 & \mathrm{\mathrm{[\mathrm{\mathbf{B}_{T}}]_{\mathit{i,j}}}\in\left\{ 0,1\right\} ,\,\forall\mathit{i,j},}\label{bt}\\
 & \mathrm{\mathrm{[\mathrm{\mathbf{B}_{R}}]_{\mathit{i,j}}}\in\left\{ 0,1\right\} ,\,\forall\mathit{i,j},}\label{br}
\end{align}
where the explicit expression for the objective, $\mathit{P}_{\mathrm{out}}^{\mathrm{DCC}}(\mathbf{p}_{\mathrm{T}},\mathbf{B}_{\mathrm{T}},\mathrm{\mathbf{B}_{\mathrm{R}}})$,
with regard to $\mathbf{p}_{\mathrm{T}}$, $\mathbf{B}_{\mathrm{T}}$,
and $\mathbf{B}_{\mathrm{R}}$ can be found by \eqref{eq: current ip},
\eqref{eq: wt}, \eqref{eq:wr}, \eqref{eq:vout}, \eqref{eq: average ith term},
and \eqref{eq:pdcc}.

As the problem has multiple variables which are highly coupled, we
adopt an alternating optimization approach to decompose it into two
sub-problems including transmit beamforming and binary antenna coding
designs.

\subsubsection{Transmit Beamforming Design}

\begin{figure*}[tbh]
\normalsize
\setcounter{MYtempeqncnt}{\value{equation}}
\setcounter{equation}{34}
\begin{equation}
\label{eq:a1}
\mathbf{a}^{(k-1)} =\frac{1}{\mathit{R}_{\mathrm{L}}}\mathrm{\left(\sum_{\mathit{n}=1}^{\mathit{N}}\sum_{\mathit{i}\,\mathrm{even},\,\mathit{i}\geq2}^{\mathit{n_{0}}}\sum_{\mathit{j}\,\mathrm{even},\,\mathit{j}\geq2}^{\mathit{n_{0}}}\mathit{\beta_{\mathit{i}}\beta_{j}\zeta_{i}\zeta_{j}}(\mathit{i+j})[\mathit{r}_{\mathit{n}}^{\left(\mathit{k-\mathrm{1}}\right)}]^{\frac{\mathit{i+j}}{2}-1}\mathbf{h}_{\mathit{n}}^{\mathit{H}}\mathbf{h}_{\mathit{n}}\right)\mathbf{p}_{T}^{(\mathit{k-1})}}.
\end{equation}
\setcounter{equation}{\value{MYtempeqncnt}}
\hrulefill
\vspace*{4pt}
\end{figure*}

Given fixed antenna coders $\mathrm{\mathbf{B}_{T}}$ and $\mathrm{\mathbf{B}_{\mathrm{R}}}$,
the channel $\mathrm{\mathbf{H}\left(\mathrm{\mathbf{B}_{T}},\mathrm{\mathbf{B}_{\mathrm{R}}}\right)}$
is fixed and $\mathit{P}_{\mathrm{out}}^{\mathrm{DCC}}$ is solely
the function of transmit beamformer $\mathbf{p}_{\mathrm{T}}$, so
that we can equivalently transform problem \eqref{p1}-\eqref{br}
into 
\begin{align}
\mathrm{\underset{\mathbf{p}_{T}}{\mathrm{max}}}\;\; & \mathrm{\mathit{P}_{\mathrm{out}}^{\mathrm{DCC}}(\mathbf{p}_{\mathrm{T}})}\label{dcsca1}\\
\mathrm{s.t.}\;\;\; & \frac{1}{2}\left\Vert \mathbf{p}_{\mathrm{T}}\right\Vert ^{2}\mathrm{\leq\mathit{P}},\label{dcsca2}
\end{align}
which is non-convex. Previous works have utilized quasi-Newton \cite{Chen2512:Antenna}
or semidefinite relaxation (SDR) methods \cite{shen2020beamforming}
to solve the problem \eqref{dcsca1} and \eqref{dcsca2}, but with
high complexity. To address this limitation, in this work we propose
a computationally efficient algorithm based on successive convex approximation
(SCA). Specifically, we introduce auxiliary variables $\mathrm{\mathrm{\mathit{r}_{\mathit{n}}=|\mathbf{h}_{\mathrm{\mathit{n}}}}\mathbf{p}_{T}|^{2}\:\forall\mathit{n}}$,
where $\mathbf{h}_{\mathrm{\mathit{n}}}=[\mathbf{H}(\mathbf{B}_{\mathrm{T}},\mathbf{B}_{\mathrm{R}})]_{n,:}$
and $\mathrm{\mathit{r}_{\mathit{n}}>0}$. According to \eqref{eq:vout}-\eqref{eq:pdcc},
the output DC power with DCC in \eqref{dcsca1} can be equivalently
rewritten as
\begin{align}
\mathit{P}_{\mathrm{out}}^{\mathrm{DCC}}=\mathrm{\frac{1}{\mathit{R}_{L}}}\sum_{n=1}^{N}\left(\mathrm{\sum_{\mathit{i}\,\mathrm{even},\mathit{i}\geq2}^{\mathit{n_{\mathrm{0}}}}\sum_{\mathit{j}\,\mathrm{even},\mathit{j}\geq2}^{\mathit{n_{\mathrm{0}}}}\mathit{\beta_{\mathit{i}}\beta_{j}\zeta_{i}\zeta_{j}}\mathit{r}_{\mathit{n}}^{\mathit{\frac{i+j}{\mathrm{2}}}}}\right).\label{eq:pdcc-appro}
\end{align}
At the $k$th iteration, using the first-order Taylor expansion, we
can approximate $\mathit{r}_{\mathit{n}}^{\mathit{\frac{i+j}{\mathrm{2}}}}$
by
\begin{align}
\mathrm{\mathit{r}_{\mathit{n}}^{\frac{\mathit{i+j}}{2}}\geq\frac{\mathit{i+j}}{2}\left(\mathit{r}_{\mathit{n}}^{\left(\mathit{k-1}\right)}\right)^{\frac{\mathit{i}+\mathit{j}}{2}-1}\mathit{r}_{\mathit{n}}-\frac{\mathit{i+j}-2}{2}\left(\mathit{r}_{\mathit{n}}^{\left(\mathit{k-1}\right)}\right)^{\frac{\mathit{i+j}}{2}}},\label{rn-1}
\end{align}
where $\mathrm{\mathit{r}_{\mathit{n}}^{\left(\mathit{k}-1\right)}}$
is the value of $\mathit{r}_{\mathit{n}}$ optimized at the $(k-1)$th
iteration and is given by $\mathrm{\mathit{r}_{\mathit{n}}^{\left(\mathit{k}-1\right)}=\mathbf{p}_{T}^{\mathit{\left(k-1\right)H}}\mathbf{h}_{\mathit{n}}^{\mathit{H}}\mathbf{h}_{\mathit{n}}\mathbf{p}_{T}^{\left(\mathit{k-}1\right)}}$.
Moreover, $\mathit{r}_{\mathit{n}}$ can be approximated by
\begin{equation}
\mathrm{\mathit{r}_{\mathit{n}}=\mathbf{p}_{T}^{\mathit{H}}\mathbf{h}_{\mathit{n}}^{\mathit{H}}\mathbf{h}_{\mathit{n}}\mathbf{p}_{T}\geq2\mathfrak{R}\left\{ \mathbf{p}_{T}^{\mathit{\left(k-1\right)H}}\mathbf{h}_{\mathit{n}}^{\mathit{H}}\mathbf{h}_{\mathit{n}}\mathbf{p}_{T}\right\} -\mathit{r}_{\mathit{n}}^{\left(\mathit{k}-1\right)}},\label{rn}
\end{equation}
where $\mathbf{p}_{\mathrm{T}}^{\left(k-1\right)}$ is the value of
$\mathbf{p}_{\mathrm{T}}$ optimized at the $(k-1)$th iteration.
Substituting \eqref{rn} into \eqref{rn-1} yields 
\begin{equation}
\mathrm{\begin{aligned}\mathrm{\mathrm{\mathit{r}_{\mathit{n}}^{\frac{\mathit{i+j}}{2}}}} & \geq\mathrm{\mathfrak{R}}\left\{ (\mathit{i+j})\left(\mathit{r}_{\mathit{n}}^{\left(\mathit{k-1}\right)}\right)^{\frac{\mathit{i+j}}{2}-1}\mathbf{p}_{\mathrm{T}}^{\mathit{\left(k-\mathrm{1}\right)H}}\mathbf{h}_{\mathit{n}}^{\mathit{H}}\mathbf{h}_{\mathit{n}}\mathbf{p}_{\mathrm{T}}\right\} \\
 & -(\mathit{i+j}-1)\left(\mathit{r}_{\mathit{n}}^{\left(\mathit{k-\mathrm{1}}\right)}\right)^{\frac{\mathit{i+j}}{2}}.
\end{aligned}
}\label{eq:rnij}
\end{equation}

Leveraging the approximation in \eqref{eq:rnij} and ignoring the
constant term, we can equivalently write the approximate problem at
the $k$th iteration as 
\begin{align}
\mathrm{\underset{\mathbf{p}_{T}}{\mathrm{max}}}\;\; & \mathrm{\mathrm{\mathfrak{R}}\left\{ \mathbf{a}^{(\mathit{k}-1)\mathit{H}}\mathbf{p}_{T}\right\} }\label{sca1}\\
\mathrm{s.t.}\;\;\; & \frac{1}{2}\left\Vert \mathbf{p}_{\mathrm{T}}\right\Vert ^{2}\mathrm{\leq\mathit{P}},\label{sca2}
\end{align}
where $\mathbf{a}^{(\mathit{k}-1)}$ is a constant given by (35).
Utilizing the Cauchy-Schwartz inequality 
\begin{equation}
\mathrm{\mathrm{\mathrm{\mathfrak{R}}}\left\{ \mathbf{a}^{(\mathit{k}-1)\mathit{H}}\mathbf{p}_{T}\right\} \leq\left|\mathbf{a}^{(\mathit{k}-1)\mathit{H}}\mathbf{p}_{T}\right|\leq\parallel\mathbf{a}^{(\mathit{k}-1)}\parallel\parallel\mathbf{p}_{T}\parallel},
\end{equation}
we can find the closed-form optimal transmit beamformer at iteration
$k$, given by 
\begin{equation}
\mathrm{\mathbf{p}_{T}^{(\mathit{k})}=\sqrt{2\mathit{P}}\frac{\mathbf{a}^{(\mathit{k}-1)}}{\parallel\mathbf{a}^{(\mathit{k}-1)}\parallel}}.\label{close}
\end{equation}
By iteratively solving the approximate problem through the closed-form
solution \eqref{close}, the proposed SCA algorithm converges to a
stationary point of the original problem \eqref{dcsca1}-\eqref{dcsca2}.
Compared with previous methods, the proposed SCA algorithm greatly
reduces the computational complexity of transmit beamforming design.

\subsubsection{Binary Antenna Coding Design}

Given the fixed transmit beamformer $\mathbf{p}_{\mathrm{T}}$, $\mathit{P}_{\mathrm{out}}^{\mathrm{DCC}}$
is solely the function of antenna coders $\mathbf{B}_{\mathrm{T}}$
and $\mathbf{B}_{\mathrm{R}}$, so that we can equivalently transform
problem \eqref{p1}-\eqref{br} into 
\begin{align}
\mathrm{\underset{\mathbf{B}_{T},\mathbf{B}_{R}}{\mathrm{max}}\;\;} & \mathrm{\mathrm{\mathit{P}_{\mathrm{out}}^{\mathrm{DCC}}(\mathbf{B}_{\mathrm{T}},\mathrm{\mathbf{B}_{\mathrm{R}}})}}\label{eq:binary1}\\
\mathrm{s.t.}\;\;\;\; & \mathrm{\mathrm{\mathrm{[\mathrm{\mathbf{B}_{T}}]_{\mathit{i,j}}}}\in\{0,1\},\,\forall\mathit{i,j},}\label{eq:binary2}\\
 & \mathrm{\mathrm{[\mathrm{\mathbf{B}_{R}}]_{\mathit{i,j}}}\in\{0,1\},\,\forall\mathit{i,j}.}\label{eq:binary3}
\end{align}
This is an NP-hard binary optimization problem that cannot be solved
by gradient-based algorithms but can be solved by heuristic searching
methods such as successive exhaustive Boolean optimization (SEBO)
\cite{shen2016successive}. The SEBO algorithm contains two stages:
1) cyclic exhaustive search over each block (of size $J$) of the
antenna coders and 2) random bit flips to explore potential better
local optimum solutions. Herein, we use a warm-start SEBO where SEBO
algorithm is run for multiple rounds and the best solution found in
one round serves as the initial point for the next round to enhance
the possibility of finding a high-quality solution. Thus, the computational
complexity of warm-start SEBO is given by $\mathcal{O}(I2^{J}W)$,
where $I$ denotes the number of iterations and $W$ denotes the number
of rounds for warm-start SEBO. The overall joint transmit beamforming
and binary antenna coding design is summarized in Algorithm 1.

\begin{algorithm}[t]
\begin{algorithmic}[1]
        \STATE \textbf{Initialize:} $i = 0$, initial transmit beamformer $\mathbf{p}^{\mathrm{(0)}}_{\mathrm{T}}$, initial antenna coders $\mathbf{B}_{\mathrm{T}}^{(0)}$, $\mathbf{B}_{\mathrm{R}}^{(0)}$, convergence threshold $\epsilon$, maximum iterations $i_{\max}$;
        \REPEAT
            \STATE $i \gets i + 1$;
            \STATE \textbf{Update Channel:} Compute the beamspace channel $\mathbf{H}(\mathbf{B}_{\mathrm{T}}^{(i-1)},\mathbf{B}_{\mathrm{R}}^{(i-1)})=\mathbf{W_{\mathit{\mathrm{R}}}^{\mathit{\mathrm{\mathit{H}}}}}(\mathbf{B}_{\mathrm{R}}^{(i-1)})\mathbf{H}_{\mathrm{C}}\mathbf{W}_{\mathrm{T}}(\mathbf{B}_{\mathrm{T}}^{(i-1)})$;
            \STATE \textbf{Optimize Transmit Beamformer $\mathbf{p}_\mathrm{T}$:}
            \STATE \quad Fix $\mathbf{B}_{\mathrm{T}}^{(i-1)}$, $\mathbf{B}_{\mathrm{R}}^{(i-1)}$;
            \STATE \quad Use SCA to solve \eqref{dcsca1} and \eqref{dcsca2} and obtain $\mathbf{p}^{(i)}_{\mathrm{T}}$;
            \STATE \textbf{Optimize Antenna Coder $( \mathbf{B}_{\mathrm{T}}, \mathbf{B}_{\mathrm{R}})$:}
            \STATE \quad Fix $\mathbf{p}^{(i)}_{\mathrm{T}}$;
            \STATE \quad Obtain $\mathbf{B}_{\mathrm{T}}^{(i)}, \mathbf{B}_{\mathrm{R}}^{(i)}$ by solving \eqref{eq:binary1}--\eqref{eq:binary3};
            
        \UNTIL{objective change $< \epsilon$ or $i = i_{\max}$}
        \STATE \textbf{Output:} Optimal $\mathbf{p}^{\star} = \mathbf{p}^{(i)}$, $\mathbf{B}_{\mathrm{T}}^{\star} = \mathbf{B}_{\mathrm{T}}^{(i)}$, $\mathbf{B}_{\mathrm{R}}^{\star} = \mathbf{B}_{\mathrm{R}}^{(i)}$.
    \end{algorithmic}

\caption{Joint Transmit Beamforming and Binary Antenna Coding Design With DCC}
\end{algorithm}

\subsection{Continuous Antenna Coding Optimization with DCC}

For continuous antenna coding based on variable reactive loads, we
formulate the joint continuous antenna coding and beamforming design
problem as
\begin{align}
\mathrm{\underset{\mathbf{p}_{T},\mathbf{B}_{T},\mathbf{B}_{R}}{\mathrm{max}}\;} & \mathrm{\mathit{P}_{\mathrm{out}}^{\mathrm{DCC}}(\mathbf{p}_{\mathrm{T}},\mathbf{B}_{\mathrm{T}},\mathrm{\mathbf{B}_{\mathrm{R}}})}\label{cont1}\\
\mathrm{s.t.}\;\;\;\; & \frac{1}{2}\left\Vert \mathbf{p}_{\mathrm{T}}\right\Vert ^{2}\mathrm{\leq\mathit{P}},\\
 & \mathrm{\mathrm{[\mathrm{\mathbf{B}_{T}}]_{\mathit{i,j}}}\in[0,1],\,\forall\mathit{i,j},}\label{eq:cont1}\\
 & \mathrm{\mathrm{[\mathrm{\mathbf{B}_{R}}]_{\mathit{i,j}}}\in[0,1],\,\forall\mathit{i,j},}\label{eq:cont2}
\end{align}
where \eqref{eq:cont1} and \eqref{eq:cont2} denote the continuous
antenna coder constraints instead. Similarly, the problems \eqref{cont1}-\eqref{eq:cont2}
can be decoupled into two sub-problems including transmit beamforming
and continuous antenna coding designs.

\subsubsection{Transmit Beamforming Design}

Given fixed antenna coders $\mathrm{\mathbf{B}_{T}}$ and $\mathrm{\mathbf{B}_{\mathrm{R}}}$,
the transmit beamforming design problem remains the same as the problem
\eqref{dcsca1} and \eqref{dcsca2}, which can be solved using the
aforementioned SCA algorithm.

\subsubsection{Continuous Antenna Coding Design}

Given the fixed transmit beamformer $\mathrm{\mathbf{p}_{T}}$, the
continuous antenna coding optimization problem is given by
\begin{align}
\mathrm{\underset{\mathbf{B}_{T},\mathbf{B}_{R}}{\mathrm{max}}\;\;} & \mathrm{\mathit{P}_{\mathrm{out}}^{\mathrm{DCC}}(\mathbf{B}_{\mathrm{T}},\mathrm{\mathbf{B}_{\mathrm{R}}})}\label{cont_coder1}\\
\mathrm{s.t.}\;\;\;\; & \mathrm{\mathrm{[\mathrm{\mathbf{B}_{T}}]_{\mathit{i,j}}}\in[0,1],\,\forall\mathit{i,j},}\\
 & \mathrm{[\mathrm{\mathbf{B}_{R}}]_{\mathit{i,j}}}\in[0,1],\,\forall\mathit{i,j},\label{cont_coder3}
\end{align}
where the antenna coder $\mathbf{B}_{\mathrm{T}}$ and $\mathbf{B}_{\mathrm{R}}$
is linear in the range {[}0, 1{]}. From \eqref{mapping-1-1}, we have
that 
\begin{align}
\left|x_{\mathrm{T},i,j}\right| & =x_{\mathrm{oc}}[\mathrm{\mathbf{B}_{T}}]_{\mathit{i,j}},\label{transform}\\
\left|x_{\mathrm{R},i,j}\right| & =x_{\mathrm{oc}}[\mathrm{\mathbf{B}_{R}}]_{\mathit{i,j}},\label{eq:transform1}
\end{align}
where $x_{\mathrm{T},i,j}$ and $x_{\mathrm{R},i,j}$ denotes the
load reactance of the $i$th variable reactive load of the $j$th
transmit and receive pixel antennas, respectively. Note that $x_{\mathrm{oc}}$
is a very large value to numerically approximate infinity and in theory
$x_{\mathrm{T},i,j}$ and $x_{\mathrm{R},i,j}$ can have arbitrary
value. Therefore, we can equivalently transform the problems \eqref{cont_coder1}-\eqref{cont_coder3}
to an unconstrained optimization problem 
\begin{align}
\mathrm{\underset{\mathbf{X}_{\mathrm{T}},\mathbf{X}_{\mathrm{R}}}{\mathrm{max}}\;\;\;} & \mathit{P}_{\mathrm{out}}^{\mathrm{DCC}}(\mathbf{X}_{\mathrm{T}},\mathbf{X}_{\mathrm{R}}),\label{eq:Continuous-1}
\end{align}
where $\mathbf{X}_{\mathrm{T}}\in\mathbb{R}^{Q\times M}$ and $\mathbf{X}_{\mathrm{R}}\in\mathbb{R}^{Q\times N}$
are constructed by $\left[\mathbf{X}_{\mathrm{T}}\right]_{i,j}=x_{\mathrm{T},i,j}$
and $\left[\mathbf{X}_{\mathrm{R}}\right]_{i,j}=x_{\mathrm{R},i,j}$
and can have arbitrary value. Therefore, we can use the Quasi-Newton
algorithm \cite{gill1972quasi} to solve the problem \eqref{eq:Continuous-1}
with convergence to a stationary point. To improve the performance,
we run the Quasi-Newton algorithm 10 times with random initial points
in the range $[-50,50]$ and subsequently select the best result.
The computational complexity of the quasi-Newton method for each iteration
is given by $\mathcal{O}((M+N)^{2}Q^{2})$ \cite{powell1987updating}.
The overall joint transmit beamforming and continuous antenna coding
design is summarized in Algorithm 2.

\begin{algorithm}[t]
\begin{algorithmic}[1]
        \STATE \textbf{Initialize:} $i = 0$, $\mathbf{p}^{\mathrm{(0)}}_{\mathrm{T}}$, $\mathbf{B}_{\mathrm{T}}^{(0)}$, $\mathbf{B}_{\mathrm{R}}^{(0)}$, $\epsilon$, $i_{\max}$;
        \REPEAT
            \STATE $i \gets i + 1$;
            \STATE \textbf{Update Channel:} Compute the beamspace channel $\mathbf{H}(\mathbf{B}_{\mathrm{T}}^{(i-1)},\mathbf{B}_{\mathrm{R}}^{(i-1)})$;
            \STATE \textbf{Optimize Transmit Beamformer $\mathbf{p}_\mathrm{T}$:}
            \STATE \quad Fix $\mathbf{B}_{\mathrm{T}}^{(i-1)}$, $\mathbf{B}_{\mathrm{R}}^{(i-1)}$;
            \STATE \quad Use SCA to solve \eqref{dcsca1} and \eqref{dcsca2} and obtain $\mathbf{p}^{(i)}_{\mathrm{T}}$;
            \STATE \textbf{Optimize Antenna Coder $( \mathbf{B}_{\mathrm{T}}, \mathbf{B}_{\mathrm{R}})$:}
            \STATE \quad Fix $\mathbf{p}^{(i)}_{\mathrm{T}}$;
            \STATE \quad Obtain $\mathbf{B}_{\mathrm{T}}^{(i)}, \mathbf{B}_{\mathrm{R}}^{(i)}$ by solving \eqref{cont_coder1}-\eqref{cont_coder3};
        \UNTIL{objective change $< \epsilon$ or $i = i_{\max}$}
        \STATE \textbf{Output:} Optimal $\mathbf{p}^{\star} = \mathbf{p}^{(i)}$, $\mathbf{B}_{\mathrm{T}}^{\star} = \mathbf{B}_{\mathrm{T}}^{(i)}$, $\mathbf{B}_{\mathrm{R}}^{\star} = \mathbf{B}_{\mathrm{R}}^{(i)}$.
    \end{algorithmic}

\caption{Joint Transmit Beamforming and Continuous Antenna Coding Design With
DCC}
\end{algorithm}

\section{Antenna Coding Optimization for MIMO WPT with RF Combining}

In this section, we formulate and solve the antenna coding optimization
problem for MIMO WPT with RF combining (RFC) to maximize the output
DC power. An efficient algorithm to jointly optimize antenna coding
and transmit and receive beamforming is also proposed.

\subsection{Antenna Coding Optimization with RFC}

For RFC scheme as shown in Fig. \ref{fig:Schematic-of-the-1}, the
signals received by the $N$ receive pixel antennas are first combined
with the receive beamformer $\mathrm{\mathbf{p}_{R}}$ to obtain the
combined received signal
\begin{equation}
\widetilde{y}(t)=\Re\{\mathbf{p}_{\mathrm{R}}^{H}\mathbf{H}(\mathbf{B}_{\mathbf{\mathrm{T}}},\mathbf{B}_{\mathrm{R}})\mathbf{p}_{\mathrm{T}}\mathit{e}^{\mathit{j\omega_{\mathrm{c}}t}}\},
\end{equation}
which is then rectified by a single rectifier. Therefore, the output
DC voltage is given by
\begin{equation}
v_{\mathrm{out}}=\sum_{\mathit{i}\,\mathrm{even},\mathit{i}\geq2}^{\mathit{n_{\mathrm{0}}}}\mathit{\beta_{\mathit{i}}\zeta_{i}}\left|\mathbf{p}_{\mathrm{R}}^{H}\mathbf{H}(\mathbf{B}_{\mathbf{\mathrm{T}}},\mathbf{B}_{\mathbf{\mathrm{R}}})\mathbf{p}_{\mathrm{T}}\right|^{i},
\end{equation}
which yields the output DC power as $\mathrm{\mathit{P}_{\mathrm{out}}^{\mathrm{RFC}}=\mathit{v}_{\mathrm{out}}^{2}/\mathit{R}_{L}}$.

We first consider binary antenna coding. We notice that $\mathit{P}_{\mathrm{out}}^{\mathrm{RFC}}$
monotonically increases with the channel gain $\left|\mathbf{p}_{\mathbf{\mathrm{R}}}^{\mathbf{\mathrm{\mathit{H}}}}\mathbf{H}(\mathbf{B}_{\mathbf{\mathrm{T}}},\mathbf{B}_{\mathbf{\mathrm{R}}})\mathbf{p}_{\mathbf{\mathrm{T}}}\right|^{2}$,
simplifying the output DC power maximization problem as
\begin{align}
\mathrm{\underset{\mathbf{p}_{\mathrm{T}},\mathbf{p}_{\mathrm{R}},\mathbf{B}_{T},\mathbf{B}_{R}}{\mathrm{max}}} & \left|\mathbf{p}_{\mathrm{R}}^{H}\mathbf{H}(\mathbf{B}_{\mathbf{\mathrm{T}}},\mathbf{B}_{\mathbf{\mathrm{R}}})\mathbf{p}_{\mathrm{T}}\right|^{2}\label{rf1-1}\\
\mathrm{s.t.}\;\;\;\;\; & \mathrm{\frac{1}{2}\left\Vert \mathbf{p}_{T}\right\Vert ^{2}\leq\mathit{P}},\\
 & \left\Vert \mathrm{\mathbf{p}_{R}}\right\Vert ^{2}\leq1,\\
 & \mathrm{\mathrm{[\mathbf{B}_{\mathbf{\mathrm{T}}}]_{\mathit{i,j}}}\in\{0,1\},\,\forall\mathit{i,j},}\label{rf4}\\
 & \mathrm{\mathrm{[\mathbf{B}_{\mathbf{\mathrm{R}}}]_{\mathit{i,j}}}\in\{0,1\},\,\forall\mathit{\mathit{i,j}}.}\label{rf5-1}
\end{align}
Given arbitrary antenna coders $\mathbf{B}_{\mathrm{T}}$ and $\mathbf{B}_{\mathrm{R}}$,
optimal transmit and receive beamforming have closed-form solutions
obtained by SVD of $\mathbf{H}(\mathbf{B}_{\mathbf{\mathrm{T}}},\mathbf{B}_{\mathbf{\mathrm{R}}})$,
written as
\begin{align}
\mathbf{p}_{\mathrm{T}}^{\star}(\mathbf{B}_{\mathbf{\mathrm{T}}},\mathbf{B}_{\mathbf{\mathrm{R}}}) & =\mathbf{v}_{1}(\mathbf{B}_{\mathbf{\mathrm{T}}},\mathbf{B}_{\mathbf{\mathrm{R}}})\sqrt{2\mathit{P}},\label{eq:pt-1}\\
\mathbf{p}_{\mathrm{R}}^{\star}(\mathbf{B}_{\mathbf{\mathrm{T}}},\mathbf{B}_{\mathbf{\mathrm{R}}}) & =\mathbf{u}_{1}(\mathbf{B}_{\mathbf{\mathrm{T}}},\mathbf{B}_{\mathbf{\mathrm{R}}}),\label{eq:pr-1}
\end{align}
where $\mathbf{v}_{1}(\mathbf{B}_{\mathbf{\mathrm{T}}},\mathbf{B}_{\mathbf{\mathrm{R}}})$
and $\mathbf{u}_{1}(\mathbf{B}_{\mathbf{\mathrm{T}}},\mathbf{B}_{\mathbf{\mathrm{R}}})$
denote the right and left singular vectors corresponding to the maximum
singular value of $\mathbf{H}(\mathbf{B}_{\mathbf{\mathrm{T}}},\mathbf{B}_{\mathbf{\mathrm{R}}})$,
denoted as $\sigma_{\mathrm{max}}\left(\mathbf{H}(\mathbf{B}_{\mathbf{\mathrm{T}}},\mathbf{B}_{\mathbf{\mathrm{R}}})\right)$.
Substituting \eqref{eq:pt-1} and \eqref{eq:pr-1} into problems \eqref{rf1-1}-\eqref{rf5-1},
we have 
\begin{align}
\mathrm{\underset{\mathbf{B}_{T},\mathbf{B}_{R}}{\mathrm{max}}\;\;} & 2\mathit{P}\left|\sigma_{\mathrm{max}}\left(\mathbf{H}(\mathbf{B}_{\mathbf{\mathrm{T}}},\mathbf{B}_{\mathbf{\mathrm{R}}})\right)\right|^{2}\label{rf1-1-1}\\
\mathrm{s.t.}\;\;\;\; & \mathrm{[\mathbf{B}_{\mathbf{\mathrm{T}}}]_{\mathit{i,j}}}\in\{0,1\},\,\forall\mathit{i,j},\\
 & \mathrm{\mathrm{[\mathbf{B}_{\mathbf{\mathrm{R}}}]_{\mathit{i,j}}}\in\{0,1\},\,\forall\mathit{\mathit{i,j}},}\label{rf5-1-1}
\end{align}
which can be solved by the warm-start SEBO algorithm.

As for the continuous antenna coding, the formulation of the output
DC power maximization problem can be easily tailored simply by replacing
\{0,1\} with {[}0,1{]} in the constraint \eqref{rf4} and \eqref{rf5-1},
so that it can be solved by the quasi-Newton algorithm similar to
problem \eqref{cont_coder1}-\eqref{cont_coder3}. For brevity, we
omit the details.

\subsection{Optimization with Analog Receive Beamforming}

\begin{figure}[t]
\begin{centering}
\includegraphics[width=8.5cm]{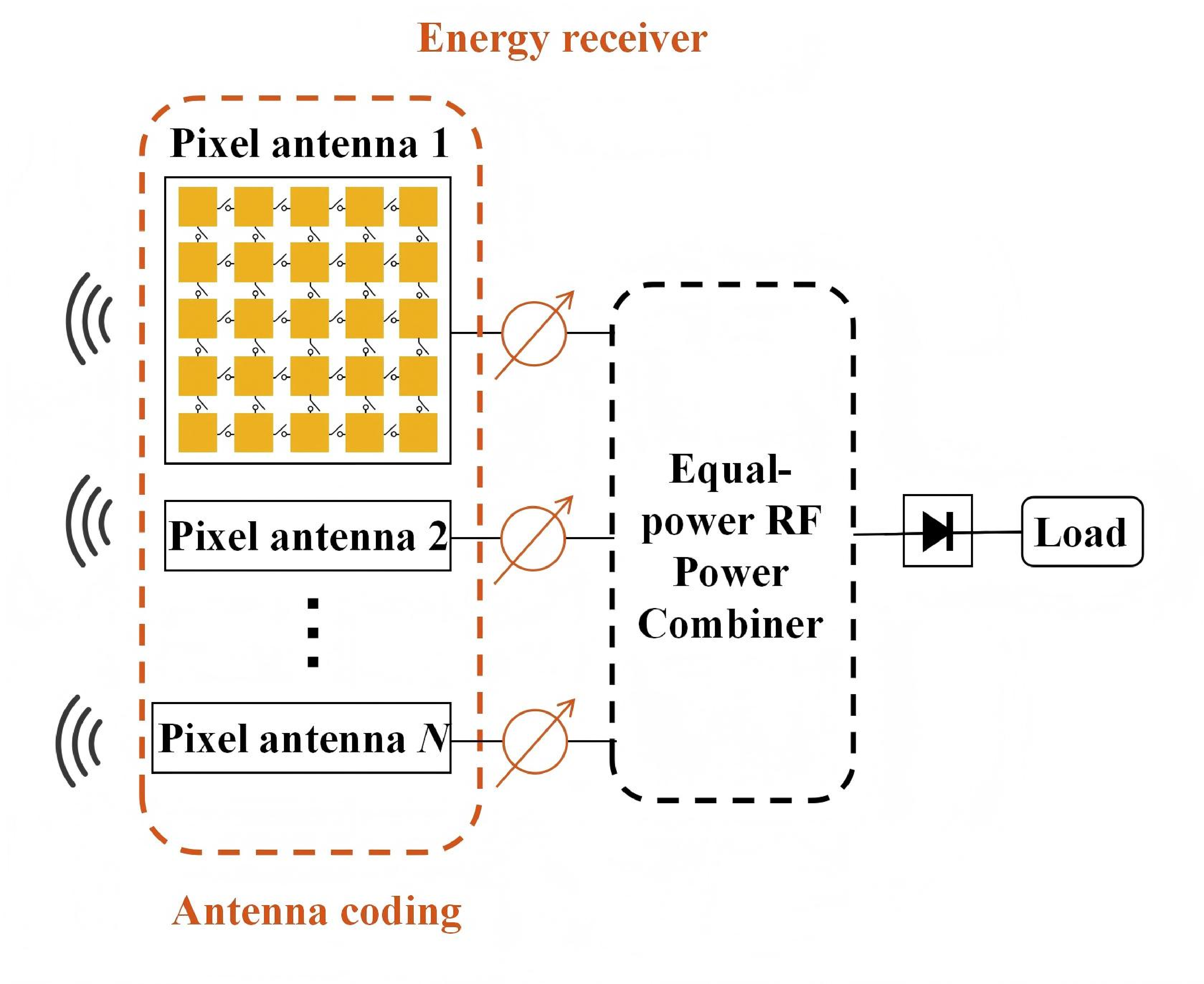}
\par\end{centering}
\caption{Schematic of the proposed energy receiver using pixel antennas with
analog receive beamforming.}\label{fig:abf}
\end{figure}

The general receive beamforming in RFC scheme requires reconfigurable
power combiners to adjust the magnitude of the receiver beamforming
weights, which however increases hardware complexity. Therefore, a
practical RFC scheme is analog receive beamforming (ABF) utilizing
phase shifters with an equal-power combiner as shown in Fig. \ref{fig:abf}.
Using the ABF, we first consider the binary antenna coding and the
output DC power maximization problem formulated as 
\begin{align}
\mathrm{\underset{\mathbf{p}_{\mathbf{\mathrm{T}}},\mathbf{p}_{\mathrm{R}},\mathbf{B}_{T},\mathbf{B}_{R}}{\mathrm{max}}\;\;} & \left|\mathbf{p}_{\mathrm{R}}^{H}\mathbf{H}(\mathbf{B}_{\mathbf{\mathrm{T}}},\mathbf{B}_{\mathbf{\mathrm{R}}})\mathbf{p}_{\mathrm{T}}\right|^{2}\\
\mathrm{s.t.}\;\;\;\;\;\;\;\;\; & \mathrm{\frac{1}{2}\left\Vert \mathbf{p}_{T}\right\Vert ^{2}\leq\mathit{P}},\\
 & \left|\left[\mathbf{p}_{\mathrm{R}}\right]_{n}\right|=\frac{1}{\sqrt{\mathrm{\mathit{N}}}},\forall n,\label{eq:modulus-1}\\
 & \mathrm{\mathrm{[\mathbf{B}_{\mathbf{\mathrm{T}}}]_{\mathit{i,j}}}\in\{0,1\},\,\forall\mathit{i,j},}\label{eq: ABF BINARY CON1}\\
 & \mathrm{\mathrm{[\mathbf{B}_{\mathbf{\mathrm{R}}}]_{\mathit{i,j}}}\in\{0,1\},\,\forall\mathit{\mathit{i,j}},}\label{eq: ABF BINARY CON2}
\end{align}
where the constant-modulus constraint \eqref{eq:modulus-1} complicates
the problem. To handle this, we adopt an alternating optimization
approach to decouple the analog receive beamforming and antenna coding
designs as follows.

\subsubsection{Analog Receive Beamforming Design}

Given fixed antenna coders $\mathbf{B}_{\mathrm{T}}$ and $\mathbf{B}_{\mathrm{R}}$,
the analog receive beamforming design problem is expressed as
\begin{align}
\underset{\mathbf{p}_{\mathbf{\mathrm{T}}},\mathbf{p}_{\mathrm{R}}}{\mathrm{max}}\;\; & \left|\mathbf{p}_{\mathrm{R}}^{H}\mathbf{H}\mathbf{p}_{\mathrm{T}}\right|^{2}\label{eq:ABF-1}\\
\mathrm{s.t.}\:\:\:\; & \frac{1}{2}\left\Vert \mathbf{p}_{\mathrm{T}}\right\Vert ^{2}\leq P,\\
 & \left|\left[\mathbf{p}_{\mathrm{R}}\right]_{n}\right|=\frac{1}{\sqrt{\mathrm{\mathit{N}}}},\forall n.\label{modulus2}
\end{align}
We notice that for any given $\mathrm{\mathbf{p}_{R}}$, the optimal
$\mathbf{p}_{\mathrm{T}}^{\star}$ that maximizes the output DC power
is given by the maximum ratio transmission (MRT), i.e. $\mathbf{p}_{\mathrm{T}}^{\star}=\mathrm{\sqrt{2\mathit{P}}}\frac{(\mathbf{p}_{\mathrm{R}}^{H}\mathbf{H})^{\mathrm{\mathit{H}}}}{\|\mathrm{\mathbf{p}_{\mathrm{R}}^{\mathit{H}}\mathbf{H}}\|}$.
Accordingly, by further introducing an auxiliary variable $\mathrm{\mathit{r}_{0}=||\mathbf{p}_{R}^{\mathit{H}}\mathbf{H}||}>0$,
the problem \eqref{eq:ABF-1}-\eqref{modulus2} can be equivalently
rewritten as 
\begin{align}
\underset{\mathbf{p}_{\mathrm{R}}}{\mathrm{max}}\;\; & \mathit{r}_{0}^{2}\label{eq:ABFSCA-1}\\
\mathrm{s.t.}\:\:\: & \left|\left[\mathbf{p}_{\mathrm{R}}\right]_{n}\right|=\frac{1}{\sqrt{\mathrm{\mathit{N}}}},\forall n.\label{eq:ABFSCA1-1}
\end{align}
To deal with the non-convex optimization problem \eqref{eq:ABFSCA-1}
and \eqref{eq:ABFSCA1-1}, we use SCA to approximate $\mathrm{\mathit{r}_{0}^{2}=\mathbf{p}_{R}^{\mathit{H}}\mathbf{HH}^{\mathit{H}}\mathbf{p}_{\mathrm{R}}}$
with its first-order Taylor expansion at the $(k-1)$th iteration,
i.e. 
\begin{equation}
\mathit{r}_{0}^{2}\geq2\mathfrak{R}\{\mathbf{p}_{\mathrm{R}}^{(k-1)H}\boldsymbol{\mathrm{HH}}^{H}\mathbf{p}_{\mathrm{R}}\}-\mathbf{p}_{\mathrm{R}}^{(k-1)H}\boldsymbol{\mathrm{HH}}^{H}\mathbf{p}_{\mathrm{R}}^{(k-1)}.
\end{equation}
Ignoring the constant term, we can formulate the approximate problem
of \eqref{eq:ABFSCA-1} and \eqref{eq:ABFSCA1-1} at $k$th iteration
as
\begin{align}
\underset{\mathbf{p}_{\mathrm{R}}}{\mathrm{max}}\;\; & \mathrm{\mathfrak{R}\{\mathbf{z}^{(\mathit{k}-1)\mathit{H}}\mathbf{p}_{\mathrm{R}}\}}\\
\mathrm{s.t.}\:\:\: & \left|\left[\mathbf{p}_{\mathrm{R}}\right]_{n}\right|=\frac{1}{\sqrt{\mathrm{\mathit{N}}}},\forall n,
\end{align}
where $\mathbf{z}^{(\mathit{k}-1)}=\mathbf{H}\mathbf{H}^{H}\mathbf{p}_{\mathrm{R}}^{(\mathit{k}-1)}$
is a constant and the closed-form optimal solution can be easily found
with 
\begin{equation}
\mathrm{\mathbf{p}_{R}^{(\mathit{k})}=\frac{1}{\sqrt{\mathrm{\mathit{N}}}}\mathit{e}^{\mathit{j}arg(\mathbf{z}^{(\mathit{k}-1)})}}.\label{eq: ABF ITERATIVE}
\end{equation}
By iteratively solving the approximate problem through the closed-form
solution \eqref{eq: ABF ITERATIVE}, the SCA algorithm converges to
a stationary point of the original problem \eqref{eq:ABF-1}-\eqref{modulus2}.
The overall analog receive beamforming design is summarized in Algorithm
3.

\begin{algorithm}[t]
\begin{algorithmic}[1]
        \STATE \textbf{Initialize:} $k = 0$, $\mathbf{p}_{\mathrm{R}}^{(0)}$, $\epsilon$, $k_{\max}$;
        \REPEAT
            \STATE $k \gets k + 1$;  
            \STATE Compute $\mathbf{z}^{(k-1)} = \mathbf{H} \mathbf{H}^H \mathbf{p}_{\mathrm{R}}^{(k-1)}$;
            \STATE Update $\mathrm{\mathbf{p}_{R}^{(\mathit{k})}=\frac{1}{\sqrt{\mathrm{\mathit{N}}}}e^{\mathit{j}arg(\mathbf{z}^{(\mathit{k}-1)})}}$;
        \UNTIL{$\|\mathbf{p}_{\mathrm{R}}^{(k)}-\mathbf{p}_{\mathrm{R}}^{(k-1)}\|/\|\mathbf{p}_{\mathrm{R}}^{(k)}\|<\epsilon $ or $k = k_{\max}$}
        \STATE Set $\mathbf{p}_{\mathrm{T}}^{(k)}=\mathrm{\sqrt{2\mathit{P}}}\frac{(\mathrm{\mathbf{p}_{R}^{(\mathit{k})\mathit{H}}\mathbf{H}})^{\mathrm{\mathit{H}}}}{\|\mathrm{\mathbf{p}_{R}^{(\mathit{k})\mathit{H}}\mathbf{H}}\|}$;
        \STATE \textbf{Output:} $\mathbf{p}_{\mathrm{T}}^{\star} = \mathbf{p}_{\mathrm{T}}^{(k)}, \mathbf{p}_{\mathrm{R}}^{\star} = \mathbf{p}_{\mathrm{R}}^{(k)}$ \unskip.
    \end{algorithmic}

\caption{Analog Receive Beamforming Design}
\end{algorithm}

\subsubsection{Antenna Coding Design}

Given fixed beamformers $\mathbf{p}_{\mathrm{T}}$ and $\mathbf{p}_{\mathrm{R}}$,
the binary antenna coding design problem can be formulated as 
\begin{align}
\mathrm{\underset{\mathbf{B}_{T},\mathbf{B}_{R}}{\mathrm{max}}\;} & \left|\mathbf{p}_{\mathrm{R}}^{H}\mathbf{H}(\mathbf{B}_{\mathbf{\mathrm{T}}},\mathbf{B}_{\mathbf{\mathrm{R}}})\mathbf{p}_{\mathrm{T}}\right|^{2}\\
\mathrm{s.t.}\;\;\; & \mathrm{[\mathbf{B}_{\mathbf{\mathrm{T}}}]_{\mathit{i,j}}}\in\{0,1\},\,\forall\mathit{i,j},\\
 & \mathrm{\mathrm{[\mathbf{B}_{\mathbf{\mathrm{R}}}]_{\mathit{i,j}}}\in\{0,1\},\,\forall\mathit{\mathit{i,j}},}
\end{align}
which can be solved by the warm-start SEBO algorithm, with $\mathbf{p}_{\mathrm{T}}$
given by MRT and fixed $\mathbf{p}_{\mathrm{R}}$.

By alternatively optimizing the transmit and receiving beamforming
and antenna coding designs, the output DC power with ABF and binary
antenna coding can be maximized. Further, the output DC power maximization
with ABF and continuous antenna coding can be easily tailored by replacing
\{0,1\} with {[}0,1{]} in the constraint \eqref{eq: ABF BINARY CON1}
and \eqref{eq: ABF BINARY CON2}, and solved by the quasi-Newton algorithm.
For brevity, we omit the details.

\section{Codebook-Based Antenna Coding Design}

To reduce the computational complexity of binary antenna coding optimization,
we propose a codebook-based antenna coding design strategy, which
is based on K-means clustering tailored for output DC power maximization
problem for the proposed MIMO WPT system.

\subsection{Problem Formulation}

A codebook design has been demonstrated in \cite{11202491} for maximizing
the average channel gain of SISO pixel antenna systems. It was based
on the generalized Lloyd algorithm (GLA), where the training channel
partitions and antenna coder centroids are alternatively optimized
until convergence. However, this method lacks consideration for MIMO
WPT configurations and the specific optimization objective beyond
channel gain. Hence, we aim to propose a scenario-oriented codebook
design method for the output DC power maximization problem with DCC
and RFC, respectively.

To this end, we consider a codebook containing $D$ different antenna
coders as $\mathcal{C}\triangleq\{\mathbf{c}_{1},...,\mathbf{c}_{D}\}$,
where $\mathbf{c}_{d}\in\mathbb{R}^{\mathit{Q\times1}}$ is the $d$th
antenna coder, also referred to as a codeword. Taking the antenna
coding optimization for MIMO WPT with RFC as an example, the output
DC power maximization problem with antenna coders selected from the
codebook $\mathcal{C}$ is written as 
\begin{align}
\mathrm{\underset{\mathbf{B}_{T},\mathbf{B}_{R}}{\mathrm{max}}\;} & \left|\sigma_{\mathrm{max}}\left(\mathbf{H}(\mathbf{B}_{\mathbf{\mathrm{T}}},\mathbf{B}_{\mathbf{\mathrm{R}}})\right)\right|^{2}\label{eq:CDB1-1}\\
\mathrm{s.t.}\:\:\:\; & \mathrm{[\mathbf{B}_{T}]_{\mathit{:,m}}}\in\mathcal{C},\,\forall\mathit{m}\mathrm{,}\label{eq:CDB2-1}\\
 & \mathrm{\mathrm{[\mathbf{B}_{R}]_{\mathit{:,n}}}\in\mathcal{C},\,\forall\mathit{n},}\label{CDB3-1}
\end{align}
where the optimal transmit and receive beamformers \eqref{eq:pt-1}
and \eqref{eq:pr-1} are implicitly utilized. However, to obtain the
optimal antenna coders $\mathrm{\mathbf{B}_{T}^{\star}}$ and $\mathrm{\mathbf{B}_{R}^{\star}}$
from the codebook, we need to first design a good codebook to maximize
the average output DC power of the MIMO WPT system, written as
\begin{align}
\mathrm{\underset{\mathcal{C}}{\mathrm{max}}\;\;} & \mathcal{E}\left\{ \left|\sigma_{\mathrm{max}}\left(\mathbf{H}(\mathbf{B}_{\mathbf{\mathrm{T}}}^{\star},\mathbf{B}_{\mathbf{\mathrm{R}}}^{\star})\right)\right|^{2}\right\} \label{eq:upper1-1}\\
\mathrm{s.t.}\:\:\: & [\mathbf{c}_{d}]_{q}\in\{0,1\},\,\forall d,q,\label{eq:upper2-1}\\
 & \mathbf{B}_{\mathbf{\mathrm{T}}}^{\star},\mathbf{B}_{\mathbf{\mathrm{R}}}^{\star}\;\mathrm{solves}\;\eqref{eq:CDB1-1}-\eqref{CDB3-1},\label{eq: upper3-1}
\end{align}
which can be regarded as an upper-level problem of \eqref{eq:CDB1-1}-\eqref{CDB3-1}.
To solve the nested optimization problems, we adopt a two-stage process:
1) the offline codebook training stage, which provides a codebook
design for antenna coding, and 2) the online deployment stage, which
searches for optimal antenna coders among the pre-designed codebook.

\subsection{Offline Codebook Training}

For the offline codebook training problem \eqref{eq:upper1-1}-\eqref{eq: upper3-1},
we assume a training set consisting of $N_{0}$ channel realizations,
following the same distribution of $\mathbf{H}_{\mathrm{C}}$
\begin{equation}
\mathcal{H}_{0}\triangleq\{\mathbf{H}_{\mathrm{C}}^{[1]},\mathbf{H}_{\mathrm{C}}^{[2]},...,\mathbf{H}_{\mathrm{C}}^{[N_{0}]}\}.
\end{equation}
Given the $N_{0}$ training channels, we first solve the output DC
power maximization problem with RFC for each training channel realization
\begin{align}
\mathrm{\underset{\mathbf{B}_{T},\mathbf{B}_{R}}{\mathrm{max}}\;\;} & \left|\sigma_{\mathrm{max}}\left(\mathbf{W}_{\mathit{\mathrm{R}}}^{\mathit{H}}\mathrm{(}\mathbf{B}_{\mathrm{R}}\mathrm{)}\mathbf{H}_{\mathrm{C}}^{[n_{0}]}\mathbf{W}_{\mathrm{T}}\mathrm{(}\mathbf{B}_{\mathrm{T}}\mathrm{)}\right)\right|^{2}\label{rf1-1-1-1}\\
\mathrm{s.t.}\;\;\; & \mathrm{[\mathbf{B}_{\mathbf{\mathrm{T}}}]_{\mathit{i,j}}}\in\{0,1\},\,\forall\mathit{i,j},\\
 & \mathrm{\mathrm{[\mathbf{B}_{\mathbf{\mathrm{R}}}]_{\mathit{i,j}}}\in\{0,1\},\,\forall\mathit{\mathit{i,j}},}\label{rf5-1-1-1}
\end{align}
where the warm-start SEBO algorithm can be used to find the optimal
antenna coders denoted as $\mathbf{B}_{\mathbf{\mathrm{T}}}^{[n_{0}]}$
and $\mathbf{B}_{\mathbf{\mathrm{R}}}^{[n_{0}]}\,\forall n_{0}$.
To facilitate the codebook design shared by all pixel antennas at
ET and ER, we collect the antenna coders for each transmit and receive
pixel antenna into a set denoted as
\begin{equation}
\mathcal{B}^{[n_{0}]}=\left\{ [\mathbf{B}_{\mathbf{\mathrm{T}}}^{[n_{0}]}]_{:,m}\mid\forall m\right\} \cup\left\{ [\mathbf{B}_{\mathbf{\mathrm{R}}}^{[n_{0}]}]_{:,n}\mid\forall n\right\} ,\label{eq:Bn0}
\end{equation}
which has $M+N$ elements. Furthermore, we collect $\mathcal{B}^{[n_{0}]}$
$\forall n_{0}$ into a set with $L=(M+N)N_{0}$ elements, that is
\begin{equation}
\mathcal{B}=\bigcup_{\forall n_{0}}\mathcal{B}^{[n_{0}]}=\{\mathbf{\bar{b}}_{1},\mathbf{\bar{b}}_{2},...,\mathbf{\bar{b}}_{L}\},\label{eq: training set of B}
\end{equation}
where $\mathbf{\bar{b}}_{l}$ denotes the $l$th element (antenna
coder). As a result, we can design the codebook $\mathcal{C}$ by
optimizing its $D$ codewords to approximate the set $\mathcal{B}$.
This is similar to a vector quantization problem, where the average
distortion between the antenna coders in the set $B$ and the codewords
is minimized in terms of Euclidean distance (i.e. Hamming distance
for binary vectors), written as
\begin{align}
\{\mathbf{c}_{1}^{\star},...,\mathbf{c}_{D}^{\star}\} & =\mathrm{\underset{[\mathbf{c}_{\mathit{d}}]_{\mathit{q}}\in\{0,1\},\forall\mathit{d},\mathit{q}}{argmin}\:\frac{1}{\mathit{L}}}\sum_{l=1}^{L}\left\Vert \mathbf{\bar{b}}_{l}-\mathbf{c}_{d}\right\Vert ,\label{quant1-1}
\end{align}
where $\mathbf{c}_{1}^{\star},...,\mathbf{c}_{D}^{\star}$ denotes
the optimized codewords.

To solve the problem \eqref{quant1-1}, which can be regarded as a
K-means clustering problem in a multi-dimensional feature space of
size $Q$, we follow an alternating optimization strategy: 1) the
antenna coder is individually assigned to their nearest codewords
as cluster centers in the current codebook in terms of Euclidean distance;
2) the cluster centers are updated to minimize the average distortion
between it and the assigned antenna coders. To characterize the antenna
coder assignment given a fixed codebook $\mathcal{C}$, we introduce
binary assignment variables $r_{l,d}\in\{0,1\}$, $l=1,...,L$, $d=1,...,D$,
which equals 1 only when the $l$th antenna coder $\mathbf{\bar{b}}_{l}$
is assigned to the $d$th codeword $\mathbf{c}_{d}$. Therefore, problem
\eqref{quant1-1} can be equivalently rewritten as
\begin{align}
\{\mathbf{c}_{1}^{\star},...,\mathbf{c}_{D}^{\star}\} & =\underset{[\mathbf{c}_{\mathit{d}}]_{\mathit{q}},r_{l,d}\in\{0,1\}}{\mathrm{argmin}}\sum_{l=1}^{L}\sum_{d=1}^{D}r_{l,d}\left\Vert \mathbf{\bar{b}}_{l}-\mathbf{c}_{d}\right\Vert ,\label{Assign-1}
\end{align}
To solve \eqref{Assign-1}, the two alternating steps are detailed
as below.

\subsubsection{Antenna Coder Assignment}

Given fixed codewords $\{\mathbf{c}_{1}^{(k-1)},...,\mathbf{c}_{D}^{(k-1)}\}$
as cluster centers, we optimize the assignment variables $r_{l,d}^{(k)}$
$\forall l,d$ at $k$th iteration by assigning $\mathbf{\bar{b}}_{l}$
to the nearest cluster center, expressed as
\begin{equation}
r_{l,d}^{(k)}=\begin{cases}
1 & \mathrm{if}\,d=\mathrm{argmin}_{i}\left\Vert \mathbf{\bar{b}}_{l}-\mathbf{c}_{i}^{(k-1)}\right\Vert \\
0 & \mathrm{else}
\end{cases},\,\forall l,d,\label{eq:assign}
\end{equation}
where $\mathbf{c}_{i}^{(k-1)}$ is the $i$th codeword (cluster center)
optimized at the $(k-1)$th iteration.

\subsubsection{Cluster Center Update}

Given fixed assignment variables $r_{l,d}^{(k)}$ $\forall l,d$,
we can update the cluster centers at $k$th iteration to minimize
the average distortion between the cluster centers and the assigned
antenna coders
\begin{align}
\{\mathbf{c}_{1}^{(k)},...,\mathbf{c}_{D}^{(k)}\} & =\mathrm{\underset{[\mathbf{c}_{\mathit{d}}^{(\mathit{k})}]_{\mathit{q}}\in\{0,1\}}{argmin}}\sum_{l=1}^{L}\sum_{d=1}^{D}r_{l,d}^{(k)}\left\Vert \mathbf{\bar{b}}_{l}-\mathbf{c}_{d}^{(k)}\right\Vert ,\label{quant1}
\end{align}
which is a binary optimization problem and can be solved by the warm-start
SEBO algorithm.

By alternatively optimizing the antenna coder assignment \eqref{eq:assign}
and cluster center update \eqref{quant1}, the average distortion
decreases over each iteration until the codebook design converges.
The overall offline codebook training is summarized in Algorithm 4.
It can be applied to different combining schemes and beamforming strategies,
and only requires an update to the objective function to obtain antenna
coder $\mathbf{\bar{b}}_{l}$ $\forall l$.

\begin{algorithm}[t]
\begin{algorithmic}[1]
        \STATE \textbf{Input:} Training channel set $\mathcal{H}_0 = \{\mathbf{H}_{\mathrm{C}}^{[1]}, \dots, \mathbf{H}_{\mathrm{C}}^{[N_0]}\}$, codebook size $D$;
        \STATE \textbf{Initialize:} $k=0,\,\mathcal{C}^{(0)}=\{\mathbf{c}_{1}^{(0)},...,\mathbf{c}_{D}^{(0)}\}$, $\epsilon$, $k_{\max}$;
        \STATE Obtain optimal antenna coder set $\mathcal{B}=\{\mathbf{\bar{b}}_{1},...,\mathbf{\bar{b}}_{L}\}$ via warm-start SEBO and \eqref{eq:Bn0}\eqref{eq: training set of B};
        \REPEAT
            \STATE \quad $k \gets k + 1$;
            \STATE \quad \textbf{Antenna Coder Assignment:} 
            \STATE \quad Fix $\{\mathbf{c}_{1}^{(k-1)},...,\mathbf{c}_{D}^{(k-1)}\}$;
            \STATE \quad Optimize $\{r_{ld}^{(k)},\,\forall l,d\}$ by \eqref{eq:assign};
            \STATE \quad \textbf{Cluster Center Update:} 
            \STATE \quad Fix $\{r_{ld}^{(k)},\,\forall l,d\}$;
            \STATE \quad Update codewords $\{\mathbf{c}_{1}^{(k)},...,\mathbf{c}_{D}^{(k)}\}$ by solving \eqref{quant1} ;
        \UNTIL{Objective change < $\epsilon$ or $k = k_{\max}$}
        \STATE \textbf{Output:} Optimized codebook $\mathcal{C}^{\star} = \{\mathbf{c}_{1}^{(k)},...,\mathbf{c}_{D}^{(k)}\}$.
\end{algorithmic}

\caption{Offline Codebook Training for Antenna Coding}
\end{algorithm}

\subsection{Online Codebook Deployment}

Once the offline codebook $\mathcal{C}$ is obtained, the antenna
coders are optimized during the online deployment stage of \eqref{eq:CDB1-1}-\eqref{CDB3-1}
with the CSI of $\mathbf{H}_{\mathrm{C}}$. This can be done by iteratively
searching the best codeword for each pixel antenna while fixing the
codewords for other pixel antennas until convergence, which is similar
to the block coordinate descent (BCD) method. During the online deployment
stage, only $(M+N)D$ searches are required for each iteration, which
is much more computationally efficient than the warm-start SEBO optimization.

\section{Performance Evaluation}

In this section, we evaluate the performance of MIMO WPT systems with
DCC and RFC with the proposed binary and continuous antenna coding
designs using pixel antennas. The codebook-based antenna coding design
is also evaluated.

\subsection{Simulation Setup and Pixel Antenna}

\begin{figure}[t]
\begin{centering}
\includegraphics[width=8.5cm]{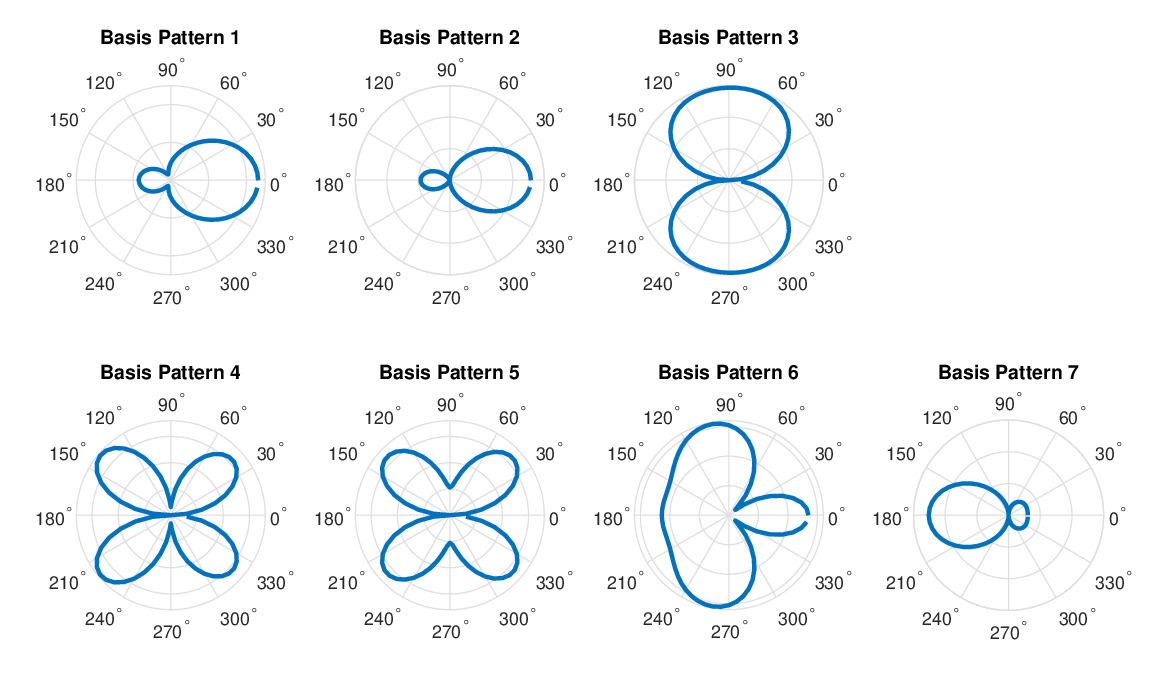}
\par\end{centering}
\caption{The orthogonal basis patterns provided by the pixel antenna in \cite{11202491}.}\label{fig:The-orthogonal-basis}
\end{figure}

A rich scattering environment is considered with a 2-D uniform power
angular spectrum and $K=72$ sampled spatial angles. The pixel antennas
deployed at ER and ET are identical but are reconfigured with different
antenna coders. Following the setting in \cite{11202491}, we consider
a $0.5\lambda\times0.5\lambda$ pixel antenna operating at 2.4 GHz
(wavelength $\lambda=125$ mm) with one antenna port and $Q=39$ pixel
ports. Obtaining the impedance matrix $\mathbf{Z}$ and open-circuit
radiation pattern matrix $\mathbf{E}_{\mathrm{oc}}$ from CST Studio
Suite, we determine the EADoF $N_{\mathrm{eff}}$ by calculating the
dominant singular values of $\mathbf{E}_{\mathrm{oc}}$ that accounts
for more than 99.8\% of total radiated power. This gives $N_{\mathrm{eff}}=7$
and the corresponding orthogonal basis patterns are shown in Fig.
\ref{fig:The-orthogonal-basis}.

Assuming a multipath Rayleigh fading environment, the entries of channel
matrix $\mathbf{H}_{\mathrm{C}}$ are modeled as i.i.d complex Gaussian
random variables, i.e. $[\mathbf{H}_{\mathrm{C}}]_{i,j}\sim\mathcal{CN}(0,1)\,\forall i,j$.
We assume 36 dBm transmit power and 66 dB path loss. The rectenna
model parameters are given as $R_{\mathrm{ant}}=50\Omega$, $R_{\mathrm{L}}=5\mathrm{k\Omega}$,
$I_{\mathrm{d}}=1.05$ , $v_{\mathrm{t}}=25\mathrm{mV}$. In the warm-start
SEBO, the block size is $J=10$ and the number of rounds is $W=20$.
The Monte Carlo method with 1000 channel realizations is utilized
to obtain the average output DC power.

\subsection{MIMO WPT Performance with Antenna Coding}

We first evaluate the performance of the proposed MIMO WPT system
utilizing pixel antennas with binary antenna coding under different
DC and RF combining schemes, including 1) the DCC with optimized (OPT)
beamforming, 2) the RFC with SVD beamforming, and 3) the RFC with
ABF. In addition, we also compare with the conventional MIMO WPT system
utilizing antennas with fixed configuration under different DC and
RF combining schemes.

\begin{figure}[t]
\centering
\includegraphics[scale=0.42]{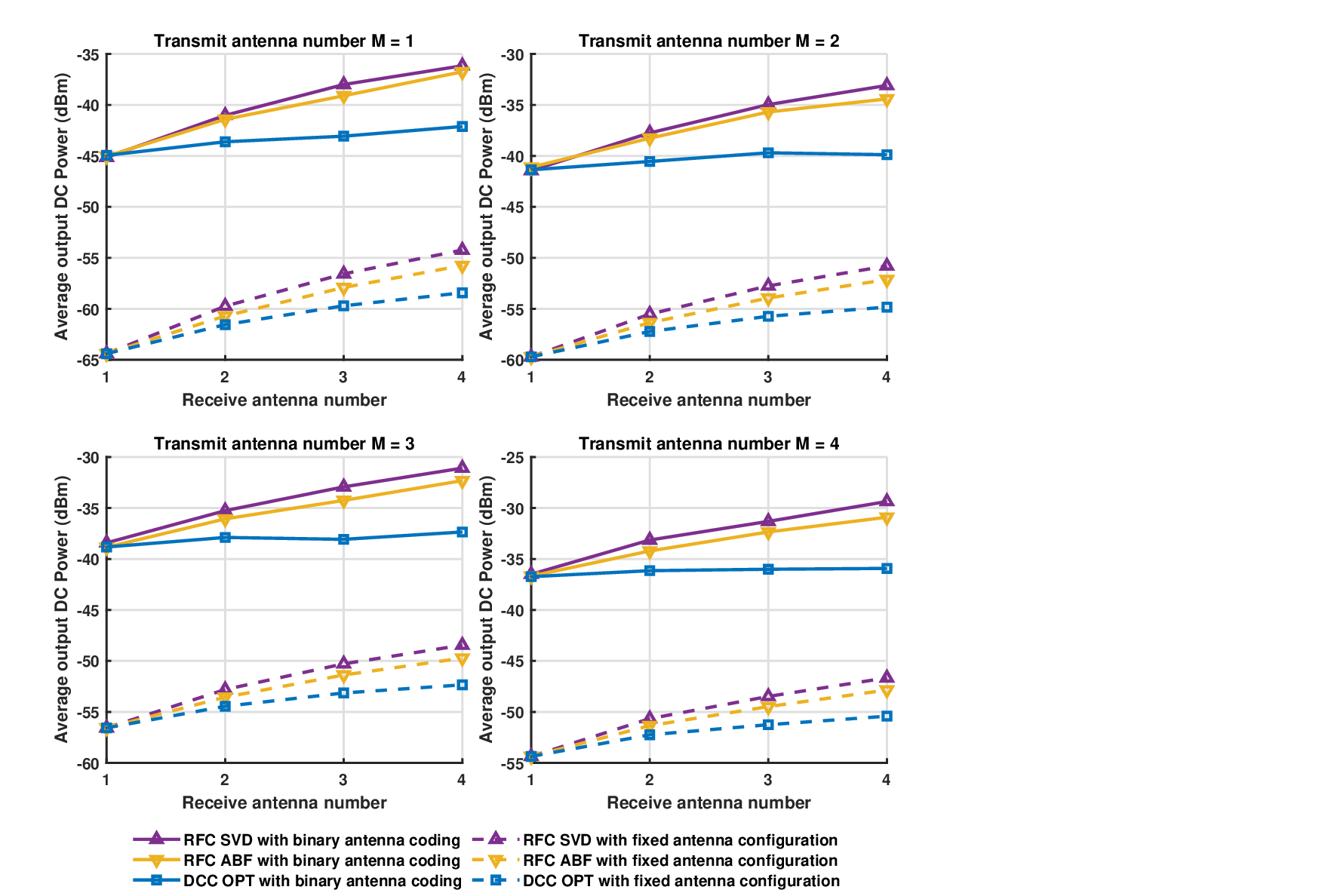} \caption{Average output DC power versus receive antenna number for different
transmit antenna numbers, with binary antenna coding and with fixed
antenna configuration, under different DC and RF combining schemes.}\label{binaryresult}
\end{figure}

Fig.~\ref{binaryresult} displays the average output DC power of
the MIMO WPT system versus the receive antenna number for different
transmit antenna numbers, with binary antenna coding and with fixed
antenna configuration, under different DC and RF combining schemes.
We can make the following observations.

\emph{First}, with the increase of transmit/receive antenna numbers,
the output DC power of MIMO WPT systems with both binary antenna coding
and fixed antenna configuration can be increased, which shows that
using pixel antennas to construct MIMO WPT systems will not affect
exploiting the spatial diversity for MIMO system.

\textit{Second}, the output DC power optimized with binary antenna
coding utilizing pixel antennas (solid lines) significantly outperforms
that with conventional fixed antenna configuration (dashed lines)
by more than 15 dB for any transmit/receive antenna number. The performance
gap becomes more prominent with increased transmit/receive antennas.
This enhancement is achieved\textbf{ }through additional degrees of
freedom introduced by antenna coding, where the radiation patterns
are flexibly reconfigured to adapt to the channel matrix, leading
to coherent adding of different multiple paths with different AoA
and AoD in the propagation environment.

\emph{Third}, RFC schemes obtain higher output DC power than DCC schemes,
with both binary antenna coding and fixed antenna configuration. This
is because the receive beamforming in RF combining can leverage the
rectenna nonlinearity more efficiently than DCC. In addition, it also
demonstrates that the extra degrees of freedom provided by pixel antennas
can synergize with exploiting the rectenna nonlinearity to jointly
enhance the output DC power in MIMO WPT systems.

\textit{Fourth}, with the constant-modulus constraint of the phase
shifter, there is some performance degradation introduced by the analog
receive beamforming compared with RFC, with both binary antenna coding
and fixed antenna configuration, but the hardware complexity can be
effectively reduced as the reconfigurable power combiner is not required.

\begin{figure}[t]
\centering
\centering{}\includegraphics[width=8.5cm]{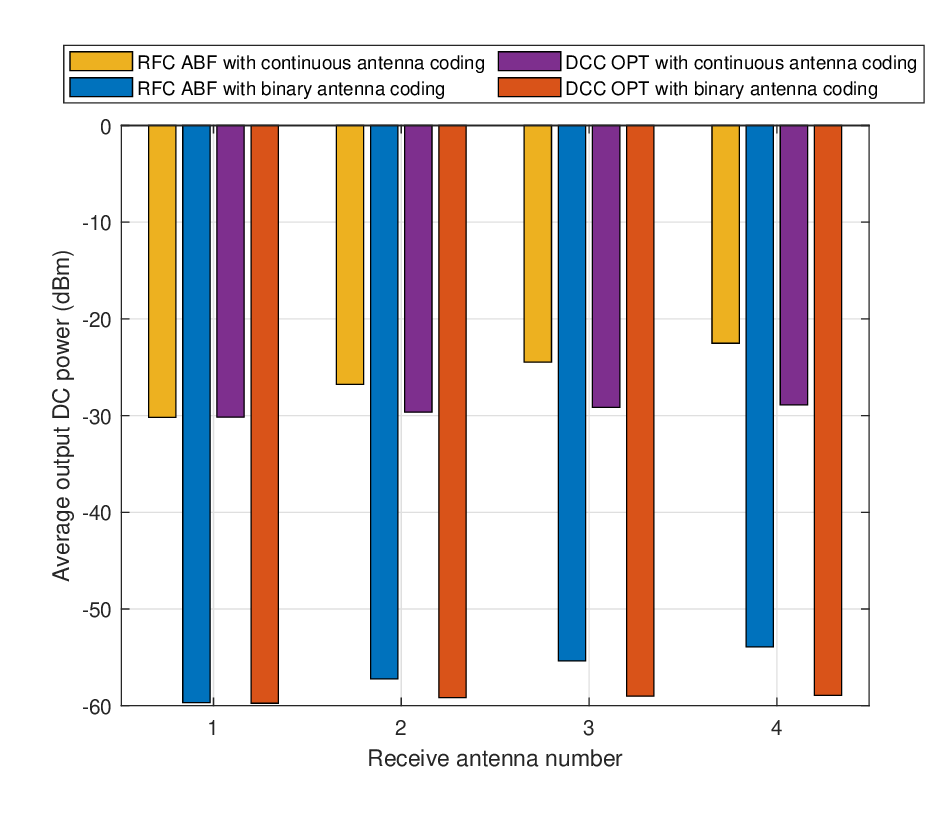}
\caption{ Average output DC power versus receive antenna number with binary
and continuous antenna coding.}\label{cont}
\end{figure}

We also evaluate the performance of the proposed MIMO WPT system with
the continuous antenna coding design in comparison with the binary
antenna coding design. The average output DC power versus receive
antenna number for $M=4$ transmit antennas, with binary and continuous
antenna coding, is shown in Fig. \ref{cont}. We can observe that
the average output DC power with continuous antenna coding outperforms
that with binary antenna coding by more than 6 dB for both RFC and
DCC schemes. This can be explained by the fact that the load reactance
of continuous antenna coding is continuous from zero to infinity while
the load reactance of binary antenna coding is either zero or infinity,
which is tighter during antenna coding optimization. Hence, continuous
antenna coding provides an upper bound of average output DC power
of MIMO WPT system using pixel antennas.

Overall, we have shown that using pixel antennas with binary and continuous
antenna coding is very beneficial for enhancing the output DC power
of MIMO WPT systems due to the extra degrees of freedom, especially
when jointly exploiting the rectenna nonlinearity to further boost
the output DC power.

\subsection{MIMO WPT Performance with Codebook Design}

We next evaluate the performance of the proposed MIMO WPT system using
pixel antennas with codebook-based antenna coding design. For the
offline codebook design, we use the $2\times2$ MIMO configuration,
which takes multiple pixel antennas into consideration while maintaining
an acceptable complexity, and $N_{0}=15000$ channel realizations
for training. We evaluate the proposed MIMO DCC/RFC OPT codebook,
which is designed by Algorithm 4 with the antenna coder training set
$\mathcal{B}$ obtained by the proposed DCC/RFC with SVD beamforming
for brevity. This is compared with benchmarks including: 1) Random
codebook, designed by random binary variable; 2) SISO channel gain
codebook, designed in \cite{11202491} to maximize the channel gain
in SISO system.

\begin{figure}
\begin{centering}
\includegraphics[width=8.5cm]{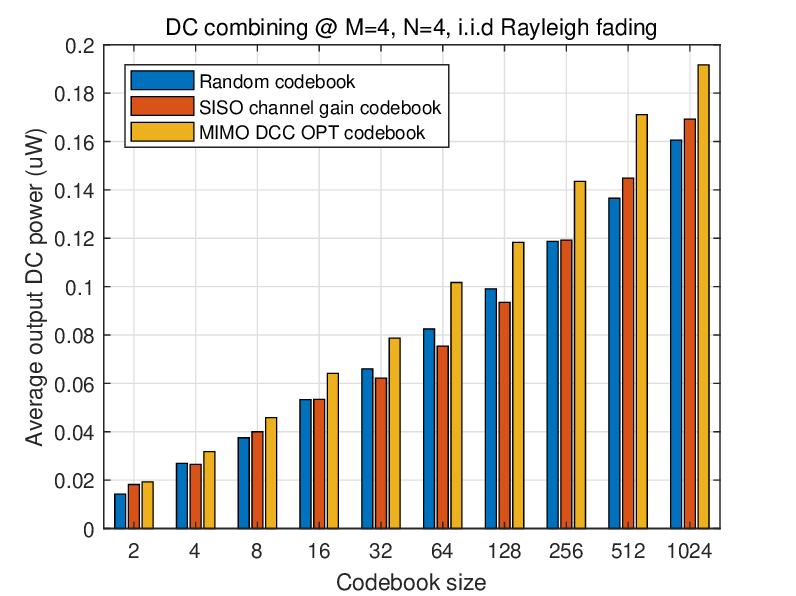}
\par\end{centering}
\caption{Average output DC power versus codebook size with codebook-based antenna
coding using DCC.}\label{fig:cdbdc}
\end{figure}

\begin{figure}
\begin{centering}
\includegraphics[width=8.5cm]{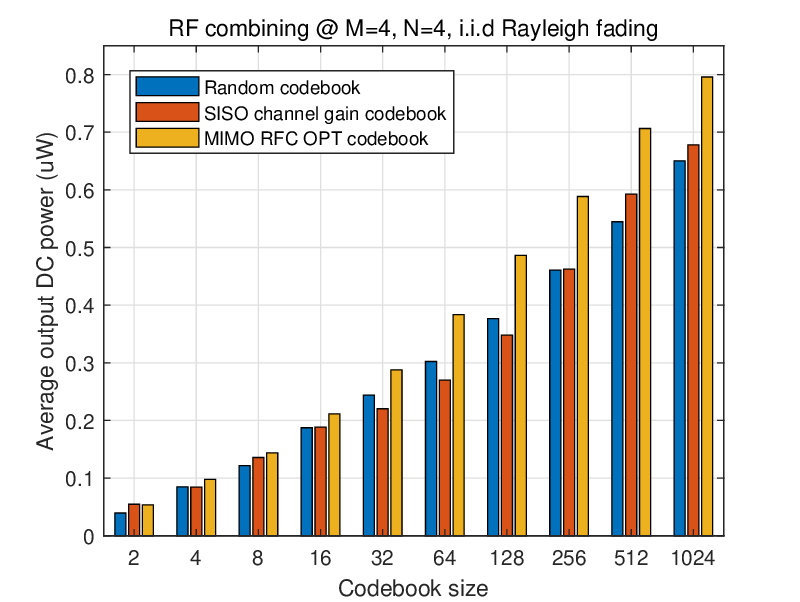}
\par\end{centering}
\caption{Average output DC power versus codebook size with codebook-based antenna
coding using RFC.}\label{fig:cdbrf}
\end{figure}

Fig. \ref{fig:cdbdc} and Fig. \ref{fig:cdbrf} display the average
output DC power of $4\times4$ MIMO WPT system versus the codebook
size with codebook-based antenna coding using DCC and RFC, deployed
online with OPT and SVD beamforming, respectively. From Fig. \ref{fig:cdbdc}
and Fig. \ref{fig:cdbrf}, we make the following observations.

\emph{First}, the average output DC power of the MIMO WPT system with
codebook-based antenna coding increases with the codebook size. This
is because increasing the codebook size enlarges the searching space
for antenna coding optimization. Therefore, the pixel antennas can
provide higher reconfigurability with more diverse radiation patterns
to adapt to the channel to enhance the MIMO WPT system. Moreover,
as the codebook size increases, the average output DC power optimized
with codebook gradually approaches that optimized with SEBO. For example,
the proposed codebook-based antenna coding design can achieve 61\%
of the output DC power optimized with SEBO for both RFC and DCC schemes,
but with significantly reduced computational complexity.

\emph{Second}, the proposed MIMO DCC/RFC OPT codebook outperforms
the random codebook and SISO channel gain codebook for all cases.
For example, in Fig. \ref{fig:cdbdc}, the proposed codebook design
can improve the output DC power by up to 35\% compared with the random
codebook. In Fig. \ref{fig:cdbrf}, the proposed codebook design can
achieve up to 40\% performance enhancement compared with the SISO
channel gain codebook. Such improvement is because the proposed codebook
design is optimized with considering the MIMO WPT configuration and
the objective of maximizing output DC power, instead of the SISO channel
gain. Therefore, it demonstrates the benefits of the proposed method
to design efficient codebook for antenna coding for MIMO WPT system.

\begin{figure}
\begin{centering}
\includegraphics[width=8.5cm]{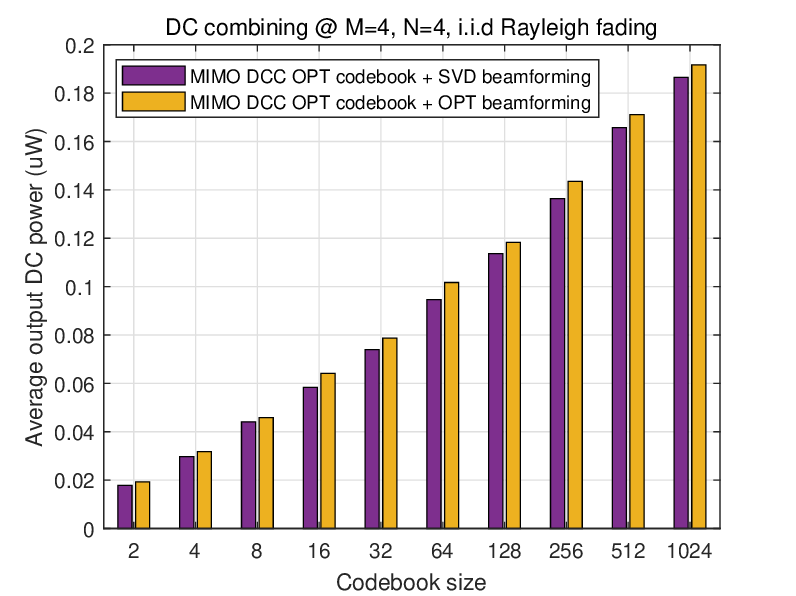}
\par\end{centering}
\caption{Average output DC power versus codebook size with codebook-based antenna
coding using DCC with different beamforming schemes.}\label{fig:cdbdc-1}
\end{figure}

To investigate the impact of rectenna nonlinearity in the proposed
pixel antenna MIMO WPT system, as shown in Fig. \ref{fig:cdbdc-1},
we consider the output DC power optimized by the MIMO DCC OPT codebook
and the aforementioned OPT beamforming. This provides a comparison
with that optimized by the MIMO DCC OPT codebook and the SVD beamforming
where the transmit beamformer is given by $\mathbf{p}_{\mathrm{T}}=\mathrm{\boldsymbol{v}_{1}\sqrt{2\mathit{P}}}$
using SVD for each channel realization that is optimal for linear
rectenna model. From Fig. \ref{fig:cdbdc-1}, we can find that the
output DC power with OPT beamforming is higher than that with SVD
beamforming. In other words, the MIMO DCC OPT codebook with SVD beamforming
is suboptimal as it overlooks the important characteristics of rectenna
nonlinearity in WPT. Hence, rectenna nonlinearity should also be properly
leveraged, together with antenna coding, to enhance the output DC
power of MIMO WPT systems utilizing pixel antennas.

Overall, the proposed codebook design achieves a good tradeoff between
output DC power and computational complexity, demonstrating its benefit
for antenna coding design for MIMO WPT systems.

\section{Conclusions}

In this work, we propose a novel antenna coding design for pixel antenna
empowered MIMO WPT systems to address the low power transfer efficiency
limitation. First, we introduce the antenna coding scheme and beamspace
channel model to demonstrate additional degrees of freedom provided
by pixel antennas. Then, we formulate the DC power maximization problem
with DC and RF combining schemes and propose joint antenna coding
and beamforming design methods, where both binary and continuous antenna
coders are considered. Efficient SCA algorithms\textbf{ }are proposed
for beamforming designs with closed-form solution updates. To reduce
the computational complexity of binary antenna coding optimization,
an efficient codebook design method based on K-means clustering is
proposed for MIMO WPT systems.

We also evaluate the proposed pixel antenna empowered MIMO WPT system
designs with antenna coding. By jointly exploiting gains from antenna
coding, beamforming, and rectenna nonlinearity, a significant\textbf{
}15 dB enhancement in average output DC power can be achieved by binary
antenna coding, compared with conventional MIMO WPT systems with fixed
antenna configuration. Moreover, the performance can be further improved
by another 6 dB with continuous antenna coding. The proposed codebook
for antenna coding also outperforms a random codebook and the previous
design by up to 40\%. It also achieves 61\% of the output DC power
optimized with SEBO binary optimization, but with much lower computational
complexity. All these results have demonstrated the potential of leveraging
antenna coding utilizing pixel antennas to enhance WPT systems.

\bibliographystyle{IEEEtran}
\bibliography{References_journal}

\end{document}